\documentclass[preprint,12pt]{elsarticle}

\usepackage{amssymb}
\usepackage{amsmath}
\usepackage{subcaption}
\usepackage[hidelinks]{hyperref}

\journal{Superconductivity}

\usepackage[T1]{fontenc}

\begin{document}


\begin{frontmatter}
\affiliation[LU]{
  organization={School of Engineering},
  addressline={Lancaster University},
  city={Lancaster},
  postcode={LA1 4YW},
  country={United Kingdom}
}

\affiliation[polito]{
  organization={Department of Applied Science and Technology},
  addressline={Politecnico di Torino},
  city={Torino},
  postcode={I-10129},
  country={Italy}
}

\affiliation[INFN]{
  organization={Istituto Nazionale di Fisica Nucleare},
  addressline={Sezione di Torino},
  city={Torino},
  postcode={I-10125},
  country={Italy}
}

\affiliation[helsinki]{
  organization={Department of Physics},
  addressline={P.O. Box 43, University of Helsinki},
  city={Helsinki},
  postcode={FI-00014},
  country={Finland}
}

\affiliation[hip]{
  organization={Helsinki Institute of Physics},
  city={Helsinki},
  country={Finland}
}

\affiliation[culham]{
  organization={United Kingdom Atomic Energy Authority},
  addressline={Culham Campus},
  city={Abingdon},
  postcode={OX14 3DB},
  country={United Kingdom}
}

\affiliation[swed]{
  organization={Theoretical Physics Division, Department of Physics, Chemistry, and Biology (IFM)},
  addressline={Link\"oping University},
  city={Link\"oping},
  postcode={58183},
  country={Sweden}
}

\affiliation[eni1]{
  organization={Eni S.p.A.},
  addressline={Piazzale Enrico Mattei, 1},
  city={Rome},
  postcode={I-00144},
  country={Italy}
}

\affiliation[eni2]{
  organization={Eni S.p.A.},
  addressline={MAFE},
  city={Venice},
  postcode={I-30175},
  country={Italy}
}

\title{Insights Into Radiation Damage in YBa$_2$Cu$_3$O$_{7-\delta}$ From Machine Learned Interatomic Potentials}
\author[LU]{Ashley Dickson\fnref{label1}}
\ead[Correspondence email address: ]{a.dickson2@lancaster.ac.uk}
\author[polito,INFN]{Niccolò Di Eugenio\fnref{label1}}

\author[polito,INFN,helsinki]{Federico Ledda}

\author[polito,INFN]{Daniele Torsello}

\author[polito,INFN]{Francesco Laviano}
    
\author[helsinki,hip]{Flyura Djurabekova}

\author[helsinki,hip]{Jesper Byggm{\"a}star}

\author[culham]{Mark R. Gilbert}

\author[culham]{Duc Nguyen-Manh}

\author[eni1]{Erik Gallo}

\author[eni2]{Antonio Trotta}

\author[swed,helsinki]{Davide Gambino}

\author[LU]{Samuel T. Murphy}

\fntext[label1]{These authors contributed equally to this work.}

\fntext[abbrv]{HTS: High Temperature Superconductors \\
YBCO: YBa$_2$Cu$_3$O$_{7-\delta}$\\
REBCO: Rare Earth Barium Copper Oxide \\
MLP: Machine Learned Potential \\
ACE: Atomic Cluster Expansion \\
tabGAP: tabulated Gaussian Approximation Potential \\
DFT: Density Functional Theory \\
TDE: Threshold Displacement Energy \\
PKA: Primary Knock-on Atom \\
MD: Molecular Dynamics \\
YBCO$_6$: YBa$_2$Cu$_3$O$_6$ \\
YBCO$_7$: YBa$_2$Cu$_3$O$_7$ \\
HRTEM: High Resolution Transmission Electron Microscope \\
GGA: Generalised Gradient Approximation \\
PBE: Perdew-Burke-Ernzerhof \\
QSD: Quasi-Static Drag \\
RDF: Radial Distribution Function\\
LAMMPS: Large-scale Atomic/Molecular Massively Parallel Simulator
}

\date{\today} 

\begin{abstract}
Accurate prediction of radiation damage in YBa$_2$Cu$_3$O$_{7-\delta}$ (YBCO) is essential for assessing the performance of high-temperature superconducting (HTS) tapes in compact fusion reactors. Existing empirical interatomic potentials have been used to model radiation damage in stoichiometric YBCO, but fail to describe oxygen-deficient compositions, which are ubiquitous in industrial Rare-Earth Barium Copper Oxide conductors and strongly influence superconducting properties.
In this work, we demonstrate that modern machine-learned interatomic potentials enable predictive modelling of radiation damage in YBCO across a broad range of oxygen stoichiometries, at a greater fidelity than previous empirical models. We employ two recently developed approaches: an Atomic Cluster Expansion (ACE) potential and a tabulated Gaussian Approximation Potential (tabGAP). Both machine-learned models are shown to accurately reproduce Density Functional Theory (DFT) energies, forces, and threshold displacement energy distributions, providing a quantitatively reliable description of atomic-scale collision processes. Molecular dynamics simulations of 5 keV cascades predict significantly enhanced peak defect production and recombination relative to a previous empirical potential, indicating qualitatively different cascade evolution. Moreover, these new machine learning models generate increased proportions of copper and oxygen vacancies compared to previous models, in direct agreement with experimental observations. By explicitly varying oxygen deficiency, we further show that total defect production exhibits only a weak dependence on stoichiometry, providing new insight into the robustness of radiation damage processes in oxygen-deficient YBCO. Finally, fusion-relevant 300 keV cascade simulations reveal amorphous regions with characteristic dimensions comparable to the superconducting coherence length, consistent with electron microscopy observations of neutron-irradiated HTS tapes. These results establish machine-learned interatomic potentials as powerful, computationally efficient tools for uncovering radiation-damage physics in YBCO, enabling predictive simulations across technologically relevant compositions and irradiation conditions.
\end{abstract}

\end{frontmatter}


\section{Introduction} \label{sec:introduction}

HTS magnets deliver high field strengths at relatively high temperatures and so offer great promise for use in compact fusion reactors \cite{torsello2025roadmap}. The functional materials in these magnets are typically the Rare Earth Barium Copper Oxide ceramics, of which YBa$_2$Cu$_3$O$_{7-\delta}$ (YBCO) is the prototypical example. During reactor operation they are subject to bombardment with high energy neutrons from the fusion plasma. Consequently, it is essential to understand how these neutrons will damage the structure of these HTS materials and how this affects their superconducting properties.\\

Neutrons transfer their energy to the lattice primarily through elastic collisions, producing high-energy PKAs that induce a sequence of atomic collisions and displacing atoms from their lattice sites to create defects, a process called a collision cascade \cite{nordlund2018primary}. These cascades proceed through three characteristic stages \cite{nordlund2018primary}. First, is the ballistic phase, during which the PKA transfers energy through a series of rapid, high-energy elastic collisions. This is followed by the heat-spike phase, where the dense cascade core rapidly rises in temperature, allowing partial local melting and subsequent recrystallization. Finally, as the region cools, the remaining primary damage may anneal out over longer timescales.\\

In the absence of a fusion-relevant neutron source, prediction of the extent and type of damage observed in YBCO tapes represents a major challenge \cite{torsello2025roadmap}. Modelling and simulation thus play a crucial role in predicting the microstructural evolution during radiation-damage cascades \cite{gray2022molecular, torsello2022expected}. The predictive power of modelling is further enhanced when complemented with experimental investigations using neutron irradiation \cite{Unterrainer_2024} and proxy techniques such as light ion irradiation \cite{sefrioui2001vortex, xiong1988transport}.\\

MD allows the simulation of primary radiation damage with atomic resolution, however, these simulations require an interatomic potential to model the interactions between atoms in the material. For YBCO in particular, development of such a potential is associated with a number of challenges: for instance, the copper environments (Cu1 and Cu2, see figure \ref{QSD}a for a structural diagram of YBCO) have differing charge states \cite{murphy2020point}, it exhibits considerable non-stoichiometry, and the layered anisotropic structure offers numerous degrees of freedom. Previous potentials for YBCO \cite{baetzold1988atomistic, chaplot1990interatomic, gray2022molecular} have been employed to understand the properties of bulk YBCO as well as its response to irradiation \cite{gray2022molecular,torsello2022expected,dickson2025threshold}. However, the Buckingham-Coulomb formalism employed by Gray \emph{et al}. cannot be used to study YBa$_2$Cu$_3$O$_{7-\delta}$ where $\delta\neq0$ as this would create an unphysical charge imbalance. Most industrially-produced HTS tapes utilise inherently sub-stoichiometric YBCO. Further, HTS tapes also accumulate additional damage over their oxygen sublattice during their operational lifetime \cite{navacerrada2000critical, swinnea1987crystal, valles1989ion, iliffe2021situ, nicholls2022understanding}. Capturing this oxygen-deficient regime is, therefore, essential for the development of an accurate and reliable model.\\

MLPs have gained significant attention in recent years due to their unique combination of flexibility, computational efficiency, and near-DFT accuracy \cite{zhao2023complex,rowe2020accurate, bartok2018machine, jafary2019applying,Eckhoff2023}. Their adaptability makes them well-suited for modelling complex materials, including systems with multiple polymorphs or variable stoichiometries \cite{zhao2023complex, rowe2020accurate, bartok2018machine}. Notably, MLPs have already demonstrated success across a wide range of material classes, from single elements \cite{byggmastar2019machine, qamar2023atomic}, to multi-element alloys \cite{jafary2019applying}, and even complex oxides \cite{zhao2023complex}. \\


In our previous work \cite{MLPYBCO} we developed 4 MLPs for YBCO. For the purposes of this work, we only employ and compare two of these potentials: the ACE \cite{drautz2019atomic} and the tabGAP \cite{bartok2010gaussian, byggmastar2021modeling} MLPs (available at \cite{DiEugenio2025MACE} and \cite{Dickson2025GAP}, respectively) to explore radiation damage in YBCO. These potentials were selected due to their computational efficiency, which allows them to used to study large scale radiation damage events. As outlined in previous work, all MLPs demonstrate considerable improvements in accuracy over the Gray potential\cite{gray2022molecular}, while also enabling simulation of non-stoichiometric YBCO.\\

In this work, we explore TDEs compared to DFT results \cite{dickson2025threshold} and to the Gray potential \cite{gray2022molecular}, investigate the curvature of the potential energy surface with respect to displacement, and validate the accuracy of our potentials for the liquid phase of YBCO (as the heat spike of a cascade is analogous to recrystallization of hot liquid \cite{nordlund2006molecular}). After this initial validation, we explore radiation damage in YBa$_2$Cu$_3$O$_{7-\delta}$ where $0\leq\delta\leq0.4$. Finally, we perform large-scale, fusion-relevant collision cascades in YBCO at an oxygen concentration typical of commercial tapes and generate HRTEM images of the resulting damage to show that the amorphous cores formed from the cascades are on the order of the coherence length, accordant with experimental observations \cite{fischer2018effect, linden2022analysing, frischherz1994defect}. \\

\section{Methods}

\subsection{Calculating forces}

This study employs a mixture of simulations employing interatomic potentials and electronic structure DFT to represent the forces between the atoms, these are detailed in this section.

\subsubsection{Interatomic potentials}

All simulations using interatomic potentials were carried out with the LAMMPS \cite{thompson2022lammps} software package. The empirical potential employed is that of Gray \emph{et al}. \cite{gray2022molecular} and the two MLPs were the tabGAP and ACE  outlined in previous work \cite{MLPYBCO}. ACE and tabGAP utilise a many-body expansion framework. In the case of tabGAP, the potential energy surface is constructed as a function of pairs, triplets and pair densities. For the ACE, a body order of up to 7 is utilised \cite{MLPYBCO}. The advantage of both frameworks is that they utilise flexible functions that are relatively unconstrained, allowing them to be fit closely to DFT data, for a wide range of atomic configurations.  Note that all ACE simulations used the performative implementation in LAMMPS \cite{lysogorskiy2021performant}. Further information on the tabGAP and ACE formalism can be found in the Appendix.  

\subsubsection{DFT simulations}

All DFT simulations were performed with CP2K \cite{hutter2014cp2k}, using the same simulation parameters as chosen for the fitting of the MLPs \cite{MLPYBCO}. We use the GGA functional framework with the PBE formalism \cite{perdew1996generalized}. Atoms are represented in a mixed Gaussian and Plane-Wave basis, from the Quickstep \cite{vandevondele2005quickstep} package in CP2K. Atomic orbitals are represented with a Double-Zeta Valence Polarised basis. Further, we use the dual-space, norm conserving, separable Godedecker-Teter-Hutter psuedopotentials optimised for PBE \cite{krack2005pseudopotentials}. A plane-wave cutoff of 400 Ry proved sufficient for energy convergence within 1 meV in total energy. All simulations were performed at the $\Gamma$ point and utilise periodic boundaries in the $x$, $y$ and $z$ directions. 

\subsection{Threshold Displacement Energies}

The TDE is the minimum amount of energy that must be transferred to an atom (the PKA) for it to produce a stable defect, either by direct collision or a chain of collisions. It can be averaged over all directions to give the average TDE ($E_{d,ave}^{av}$). This value effectively enables prediction of the radiation hardness of the material when used in analytical damage models / codes (e.g. SRIM \cite{ziegler2010srim}), therefore, accurate computation of the quantity is crucial. For direct comparison to DFT results for the TDEs in YBCO, we adopt the same methodology as Dickson \textit{et al.} \cite{dickson2025threshold}, where the $E_{d,ave}^{av}$ of the O1 atom in YBCO was calculated at 25 and 360 K using DFT-MD. These temperatures reflect the operational fusion magnet temperature and the elevated temperatures for neutron irradiation experiments in the TRIGA reactor, respectively.  \\

Simulations were performed in a $6\times6\times2$ supercell (shown to be sufficient to contain the cascades due to the low TDEs of oxygen in YBCO). The cells are equilibrated in an NPT ensemble for 5 ps, at 25 K and 360 K. During the TDE simulations, a boundary Langevin thermostat with a damping parameter of 0.1 ps is applied to the outer $2${\AA}, preventing the thermal shock interacting with its periodic image in an unphysical way, while the remaining atoms are modelled under the microcanonical ensemble (i.e. NVE). \\

At the start of this period the PKA is given an energy of 1 eV in the direction of interest. The simulations are then run for 400 fs (with a timestep of 1 fs). This time was shown in previous work to be sufficient to reach the conclusion of the displacement event. When complete the supercells were examined for defects (vacancies, interstitials, antisites) using Wiegner-Seitz analysis implemented in the python pipeline for Ovito \cite{stukowski2009visualization}. If no defect is detected the simulation starts again and the PKA energy is increased by 1 eV. This process is repeated until a defect is detected, at which point it is considered that the TDE has been reached. 

The TDE was evaluated for the same 50 directions used in \cite{dickson2025threshold}, and the value was averaged over these 50 directions to obtain $E_{d,ave}^{av}$. This is then repeated 50 times for different equilibrations, yielding a distribution of $E_{d,ave}^{av}$. $E_{d,ave}^{av}$ values are calculated by constructing a spherical Voronoi tesselation for each vector position, and using the area of the polygon for each point as a weight for averaging over all directions (this is done using SciPy \cite{virtanen2020scipy}).

\subsection{Quasi-Static Drag}

QSD simulations involve the stepwise displacement of atoms along a given displacement direction, with all other atoms fixed. At each atomic position, a single point energy computation is performed using both the DFT and the interatomic potentials. This provides information about the curvature of the potential energy surface with respect to displacement of a given atom. \\

Simulations were performed from an energy minimised 6$\times$6$\times$2 cell of YBCO$_7$. The training data for the MLPs include displacement along high symmetry directions, due to the inclusion of strained configurations \cite{MLPYBCO}. Therefore, to provide an unbiased exploration of the behaviour of the models in regions outside of the training data, random displacement vectors are picked for each symmetrically distinct atom in YBCO (these vectors are shown in Figure \ref{QSD}b). Along these displacement vectors, an atom of interest is moved in steps of 0.1 {\AA} up to a maximum distance of 3 {\AA}. The energy and forces are then recorded. The energy of the perfect cell is subtracted from the observed energies of each displacement. The same methodology is utilised for the potentials, using the DFT minimised cell as the starting point. 

\subsection{Liquid Simulations}\label{rdfmethod}

To obtain the RDFs for liquid YBCO, we used the MLPs to initially melt a $3\times3\times1$ cell of YBCO at 2000 K using NPT molecular dynamics. We then performed DFT MD on this cell for 5 ps (with a timestep of 1 fs) in an NVT ensemble. For the MLPs and the Gray potential, we generated melted cells with identical density to the DFT cell, and performed the same method with some alterations: the NVT run with each potential was carried out for 50 ps (timestep 1 fs), and the chosen cell size was $6\times6\times2$. The RDFs were averaged over all positions for every 100 timesteps for the potentials, and every 1 timestep for DFT.\\

\subsection{Cascade Simulations}

A number of different types of cascade simulations were performed using the same methodology employed for all cascades, with the exception of PKA energies, directions, and cell sizes / compositions. Smaller cascades use a cell size of around 3 million atoms, whereas the large 300 keV cascades use a cell size of 350 million atoms. All cells are first equilibrated for 20 ps (timestep 1 fs) in an NPT ensemble at 25 K as the likely operational temperature of fusion magnets (40 ps is used for the 300 keV simulations). For the lower energy cascades, the cell is then switched to a hybrid NVE/Langevin dynamics mode, where the central atoms undergo NVE MD and the border atoms (with a padding of 3 {\AA}) are attached to a Langevin thermostat, keeping the temperature at 25 K. We use a damping parameter of 0.1 ps for the boundary thermostat. For the larger 300 keV cascade, we utilise a hybrid NVE/velocity-rescaling approach (where boundary atoms in the outer 3 {\AA} have their velocities rescaled to 25 K). This is primarily due to the Langevin thermostat not being strong enough to prevent reflection of the thermal shock through the periodic boundaries for these higher energy simulations, leading to spurious defect counts. An adaptive timestep is employed for the cascade section of the simulations, with a maximum timestep of 1 fs. This ensures simulations remain stable despite high PKA velocities. To initiate the cascade, the PKA is given a velocity vector in a predetermined direction, and the cascade is allowed to evolve. Again, defect counting is performed externally with the Ovito \cite{stukowski2009visualization} Python wrapper, utilising Wigner-Seitz analysis. The cascade section of these simulations is allowed to evolve for 8 ps, with the exception of the 300 keV cascades which evolve for 40,000 timesteps (corresponding to around 20 ps, due to the usage of an adaptive timestep). At higher energies, energy loss attributed to electronic excitation is significant. Therefore, for all simulations (unless stated otherwise), we employ electronic stopping as a friction term parametrised by SRIM \cite{ziegler2010srim} for all energies above 10 eV.\\

To create simulations of sub-stoichiometric simulations, oxygen atoms are randomly removed from the chains to reach the desired stoichiometry. This is repeated for each individual simulation so the distribution of oxygen is never identical.\\

\subsection{HRTEM Reconstruction}

A lamella of thickness 40 nm (representative of the lamella thickness in \cite{linden2022analysing}) was extracted from a region of the supercell containing an amorphous zone of interest. The simulation package abTEM \cite{madsen2021abtem} was used to create simulated HRTEM images. The electron scattering potential was computed from atomic coordinates using the Lobato parametrization implemented in abTEM to produce the projected potential slices used by the multislice solver.\\

Thermal diffuse scattering and static disorder were modelled using the frozen-phonon approximation (we used 10 frozen phonons). The multislice wavefunction was propagated for each frozen-phonon configuration and the resulting intensities averaged to produce the final image.\\

The simulated incident electron probe and imaging conditions were configured to match the aberration-corrected HRTEM images from Linden et al \cite{linden2022analysing}. Key microscope parameters included an accelerating voltage of 300 KV, spherical aberration of 0.6 mm, chromatic aberration of 1.1 mm, and an energy spread of 0.65 eV. These parameters were obtained from Linden et al. \cite{linden2022analysing} and the specification sheet for the JEOL JEM-3000F FEGTEM machine. To emulate realistic experimental images, Poisson (shot) noise and detector readout noise were added.

\section{Results and Discussion}

\subsection{Quasi Static Drag}
\begin{figure*}
    \centering
    \includegraphics[width=\linewidth]{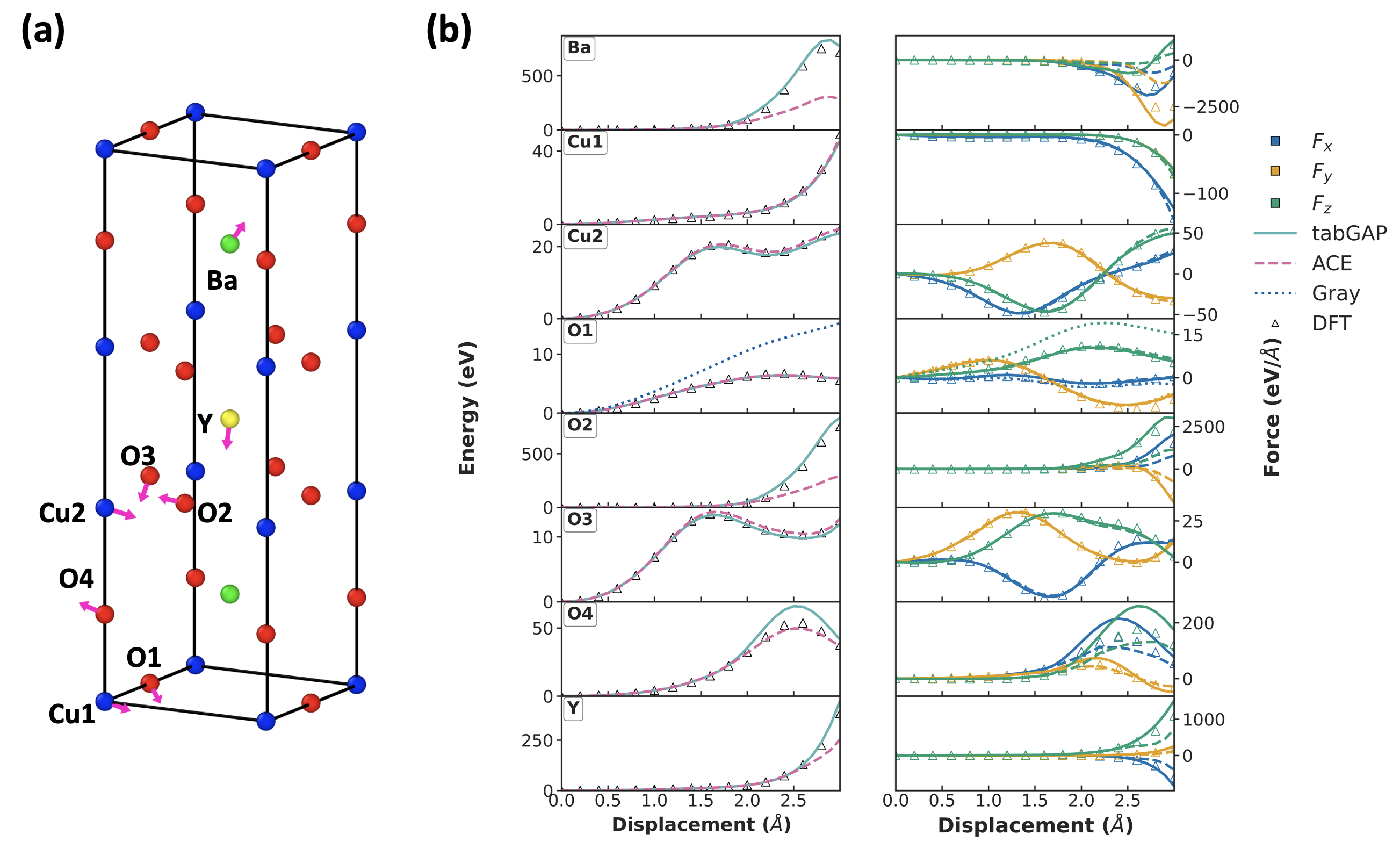}
    \caption{(a) Unit cell of YBCO with displacement vectors used for QSD overlaid as arrows in pink. Oxygen is red, barium green, yttrium yellow, and copper is blue. (b) QSD simulations, showing energies and forces (x, y and z components) for displacement of each symmetrically unique atom in YBCO. The lines are data from the potentials, and the traingles are DFT data.   }
    \label{QSD}
\end{figure*}

The energy and forces (from the displaced atom) of the corresponding displacements are shown in Figure \ref{QSD}b. For benchmarking against existing models, we also evaluate the Gray potential by reporting its energy and force predictions for the O1 atom (the chain oxygen atom) -- the primary PKA species used in the DFT TDE simulations of \cite{dickson2025threshold}. \\

In general, Figure \ref{QSD}b shows the agreement between the DFT and MLP energies is excellent. The gradients of the potential energy surface are also well reproduced, as shown by the force data in Figure \ref{QSD}b. In the higher energy regime, the tabGAP closely tracks the DFT data, however, the ACE appears to significantly underestimate the repulsive energy for Ba, O2 and Y. By contrast, the tabGAP overestimates the energy of O4 displacement, which is well captured by the ACE (although the difference here is on the order of 10 eV, whereas the ACE discrepancies are on the order of 100s of eV). These differences arise due to the current limitations of including short range interactions to ACE potentials, as discussed in the Appendix. Interestingly, as shown in sections \ref{cascades} and \ref{substoichcascades}, this does not appear to significantly affect cascades. This can be understood by considering that the Ba PKA in the first step of a typical 5 keV simulated cascade moves less than 0.1 {\AA}, so that the long distance QSD behaviour has a small probability of affecting the cascade data. \\

Nordlund \textit{et al.} \cite{nordlund2025repulsive} have demonstrated that their all electron DFT repulsive potentials (which were used for the tabGAP) outperform the standard ZBL for replicating empirical silicon ion implantation depths, thereby suggesting that the tabGAP should give improved predictions for the higher energy regime compared to the Gray potential, which uses a standard ZBL. Notably, the Gray potential shows a significant difference to the DFT data for energies and forces with respect to random displacement of the O1 atom. For the same displacement, our MLPs show significantly improved agreement to DFT data. This insight is important for the following introduction of TDEs. \\

\subsection{Threshold Displacement Energies}

Our previous work investigated TDEs in YBCO with DFT \cite{dickson2025threshold}, utilising similar simulation parameters as those used to train the MLPs. The DFT TDE was obtained by equilibrating a supercell of YBCO (at 25 and 360 K), selecting an O1 PKA in the centre of the cell, and determining the TDE in 50 different directions. This was then averaged over all directions to give the quantity $E_{d,ave}^{av}$. This can be interpreted as the \textit{average TDE over directions}, but does not reflect how thermal perturbation affects the TDEs. The ideal case, therefore, is to determine many $E_{d,ave}^{av}$ values, generating a distribution. The average of this distribution is then the \textit{average TDE over directions} AND \textit{thermal perturbations}. Here, we call this average $\langle E_{d,ave}^{av} \rangle$. It is this value which should be used as the input TDE for codes such as SRIM \cite{ziegler2010srim}. Given the computational cost of DFT, we were not able to obtain $\langle E_{d,ave}^{av} \rangle$, but we do have a single $E_{d,ave}^{av}$ (at each temperature) drawn from the ``true'' $E_{d,ave}^{av}$ distribution over thermal perturbations.\\

\begin{figure*}
    \centering
    \includegraphics[width=\linewidth]{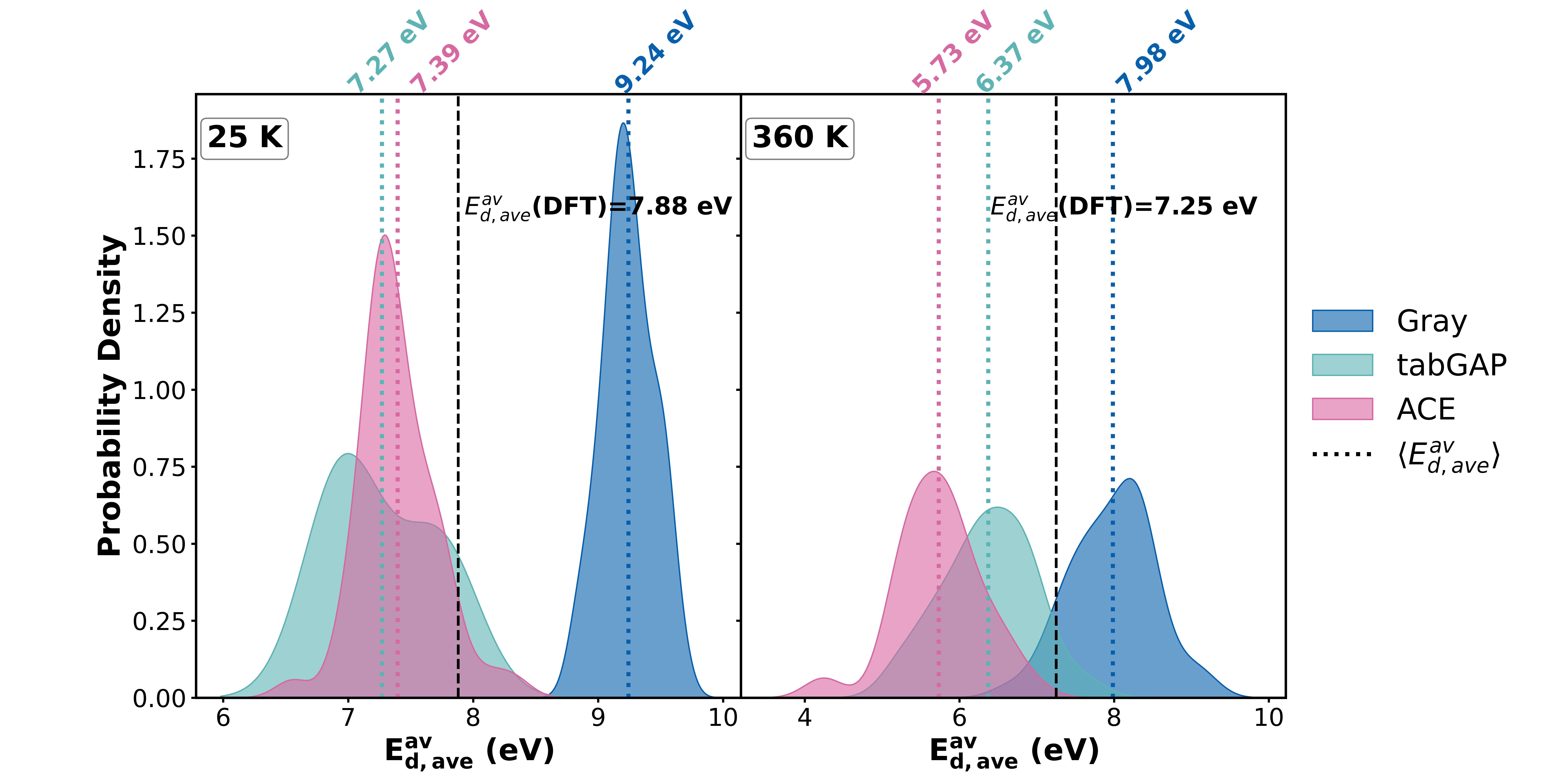}
    \caption{TDE distributions for the O1 atom at 25 K and 360 K. The distributions are each composed of 50 average TDEs ($E_{d,ave}^{av}$). Each $E_{d,ave}^{av}$ is determined from a different starting equilibration. The results are compared for the ACE, tabGAP and Gray potentials. A DFT $E_{d,ave}^{av}$ value \cite{dickson2025threshold} is also overlaid for each temperature as a dashed black line. The dotted lines in the centre of each distribution denote the average of the 50 $E_{d,ave}^{av}$ values obtained for each potential ($\langle E_{d,ave}^{av} \rangle$). }
    \label{TDE}
\end{figure*}

The MLPs (and Gray potential) can therefore be compared to DFT by generating $E_{d,ave}^{av}$ distributions at each temperature, and checking if the single DFT $E_{d,ave}^{av}$ value falls within the distribution. If this is the case, we know that a given potential is producing an $E_{d,ave}^{av}$ distribution at least similar to that of DFT. Figure \ref{TDE} shows this comparison. The $E_{d,ave}^{av}$ distribution for each potential is displayed, with the $\langle E_{d,ave}^{av} \rangle$ values marked with dotted lines. The single DFT $E_{d,ave}^{av}$ value is shown as a dashed black line. At both 25 and 360 K, this DFT value is found within the ACE and tabGAP $E_{d,ave}^{av}$  distributions, affirming that the MLPs are replicating DFT behaviour. We note that at both temperatures, the probability that the DFT $E_{d,ave}^{av}$ value was drawn from a distribution similar to the tabGAP is higher than that of the ACE. We therefore propose that the tabGAP likely gives a more accurate $E_{d,ave}^{av}$ distribution, leading to a more accurate overall $\langle E_{d,ave}^{av} \rangle$ (especially given the QSD results in Figure \ref{QSD}b). The Gray potential appears to overestimate the $E_{d,ave}^{av}$ distribution at 25 K, performing significantly worse than the MLPs. The distribution at 360 K is also shifted to higher energies, however, it does contain the DFT $E_{d,ave}^{av}$, at (surprisingly) a higher probability than the ACE. However, given the results at 25 K, we maintain that the MLPs are still likely to exhibit more accurate TDE values than the Gray potential. \\ 

Byggm{\"a}star \textit{et al.} demonstrate that the choice of short range potential gives rise to significantly different radiation damage behaviour in iron \cite{byggmastar2018effects}. The authors observed that the number of defects generated in collision cascades was directly correlated with TDEs, which themselves were dependent on the stiffness of the repulsive potential. Therefore, the more accurate short range repulsive terms of the tabGAP are expected to give more accurate TDEs than the Gray potential. However, for the energies considered here, this difference will be subtle. It is therefore reasonable to assume the more accurate potential energy surface of both the ACE and tabGAP potentials instead gives rise to the higher fidelity TDE values in Figure \ref{TDE}. \\

Despite the observed differences in the $E_{d,ave}^{av}$ distributions, the $\langle E_{d,ave}^{av} \rangle$ values for the MLPs are very similar; even at 360 K $\langle E_{d,ave}^{av} \rangle_{tabGAP}$ only differs from $\langle E_{d,ave}^{av} \rangle_{ACE}$ by around 0.5 eV, while $\langle E_{d,ave}^{av} \rangle_{Gray}$ shows a 1-2 eV difference from the MLP values. Given the small energies considered here, this difference is significant, indicating that primary damage responses for the MLPs and the Gray potential will be quite different (as seen later in this paper).

\subsection{Liquid Radial Distribution Functions}

\begin{figure*}
    \centering
    \includegraphics[width=\linewidth]{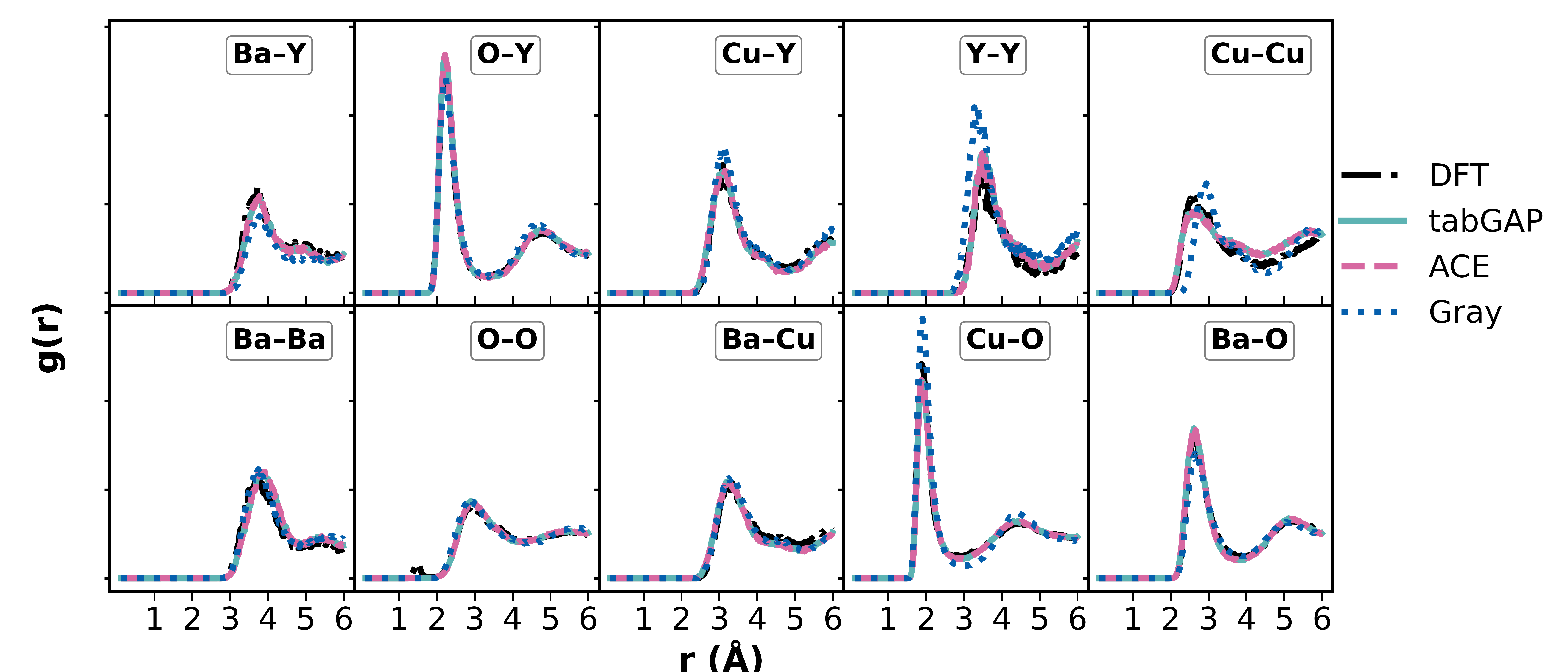}
    \caption{Radial distribution functions for each pair in YBCO at a density of 6.8 $gcm^{-3}$. }
    \label{RDF}
\end{figure*}

The RDFs for liquid YBCO at 2000 K from the DFT, MLPs and Gray potential are shown in Figure \ref{RDF}. The qualitative agreement between the DFT and MLP results is excellent. All features from the DFT RDF profiles are reproduced, with the exception of the small oxygen-oxygen peak at 1.6 {\AA}. This is a highly strained oxygen bond and likely corresponds to some form of weakly bound cuprate (like those exhibited in \cite{dickson2025threshold}). Whether or not these structures are physical is unclear, as GGA functionals have a tendency to overstabilise unrealistic peroxide-like species \cite{youssef2012intrinsic, erhart2005first, murphy2013anisotropic}. In any case, the MLPs represent an improvement over the results of the Gray potential. Given the improved RDF profiles of the MLPs, the predicted outcomes of high-energy cascades, where the morphology of the heat spike plays a critical role, are expected to be more reliable.\\

\subsection{Collision Cascades in Stoichiometric YBCO}\label{cascades}
\begin{figure*}
    \centering
    \includegraphics[width=\linewidth]{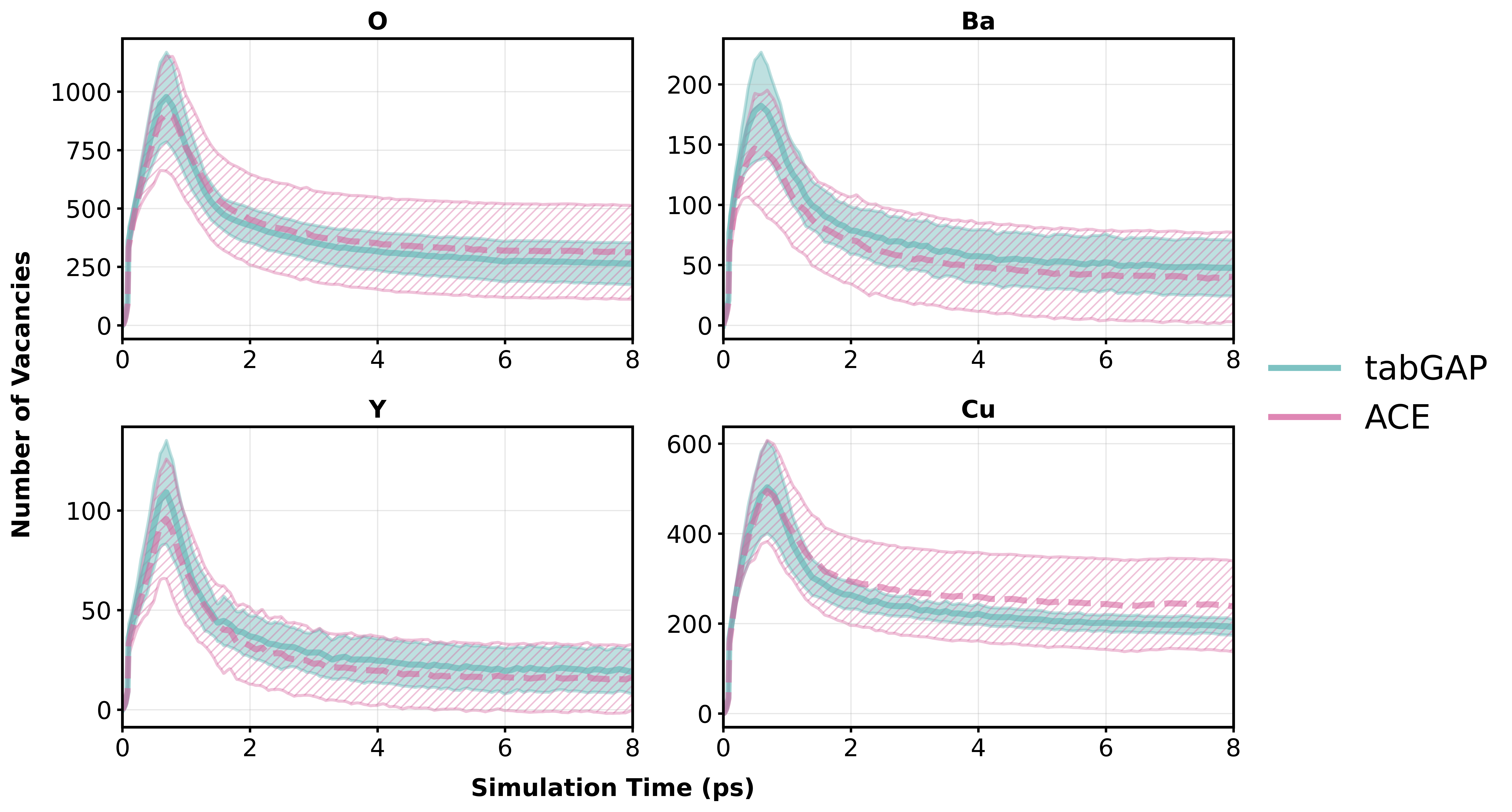}
    \caption{Number of vacancies of each element type for 5 keV Ba PKA collision cascades as a function of time. The lines are the averages of all cascades, and the shaded areas are the standard deviations. }
    \label{vacvtime}
\end{figure*}

With all the above information in mind, we can expect the MLPs to be accurate when it comes to large collision cascades. To compare the predictions of the new MLPs to the Gray potential, we simulated ten 5 keV Ba PKA collision cascades in each Cartesian direction for 8 ps (at 25 K) and averaged the resulting vacancy counts, as shown in Figure \ref{vacvtime}. Note that for direct comparison with results from Gray \textit{et al.} \cite{gray2022molecular}, we did not include electronic stopping for these simulations, and we only measured vacancy counts for defects. The averaged peak and final defect counts for each model are shown in Table \ref{defs}, with values for the Gray potential taken from collision cascades of the same energy (and the same PKA) in Ref. \cite{gray2022molecular}. \\

\begin{table}
\centering
\begin{tabular}{l|l|c c|c}
\hline
 & & \multicolumn{2}{c|}{Vacancies} &  \\
   Model   &    Element     & Peak & Final & Recombination (\%) \\
\hline
tabGAP    & O       & 977            & 264             & 73 \\
      & Cu      & 503            & 192              & 62 \\
      & Ba      & 182             & 48               & 74 \\
      & Y       & 109             & 19               & 83 \\
      & \textbf{Total} & \textbf{1771} & \textbf{521} & \textbf{71} \\
\hline
ACE & O & 905 & 312 & 66 \\
      & Cu & 494 & 238 & 52 \\
      & Ba & 147 & 40 & 73 \\
      & Y & 96 & 16 & 83 \\
      &$ \textbf{Total}$ & \textbf{1642} &\textbf{606} &\textbf{63} \\
      \hline
Gray    & O       & 384            & 319             & 17 \\
      & Cu      & 116            & 25             &  78\\
      & Ba      & 34            & 6             & 82 \\
      & Y       & 18            & 2             & 89 \\
      & \textbf{Total} & \textbf{552} & \textbf{352} & \textbf{36} \\
\hline
\end{tabular}
\caption{Peak and final defect counts of the mean lines in Figure \ref{vacvtime}, with the associated recombination percentage (percentage of defects that recombine from the peak defect count). The Gray results are taken from Ref. \cite{gray2022molecular} as an average over all cascades.}
\label{defs}
\end{table}

The region where the peak number of vacancies is observed is the heat spike of the cascade. The subsequent drop is then the recombination phase associated with recrystallisation of the damaged region. There are three primary observations of interest here: (1) Overall, the MLPs yield a final defect count roughly 50 \% higher than the Gray potential. (2) The Gray potential predicts oxygen to have the lowest recombination percentage of all elements, in direct contrast to our MLPs. The larger final defect counts may suggest a greater degree of amorphisation with the ACE and tabGAP compared to the Gray. (3) The MLPs exhibit a substantially larger proportion of copper defects than the Gray potential. This directly agrees with the proclivity of experimentally observed Cu-O divacancies in irradiated YBCO samples \cite{chudy2012point}.\\

The implications for the increased vacancy counts from the MLPs are not necessarily negative for the superconducting performance of HTS tapes. Amorphisation of local regions improves the superconducting performance (although beyond a certain saturation degradation is observed), due to the production of flux pinning centres \cite{sauerzopf1993fast, umezawa1987enhanced}. Crucially, the efficiency of these flux pinning regions is directly related to the consistency of the amorphous region size with the coherence length ($\xi$) \cite{zhang2022progress}. The size of amorphous regions predicted by the MLPs is discussed later in this paper. \\

We note that the ACE has a higher standard deviation in defect counts than the tabGAP, however, the average defect counts and recombination percentages agree very well. It is plausible this difference is due to the differing thermal properties of the MLPs, as shown in Ref. \cite{MLPYBCO}. This observation may also be explained by the differences in energies and forces seen in Figure \ref{QSD}b. Given the anisotropic nature of YBCO, we also explored the effect of irradiation direction on resulting damage. The final defect counts for irradiation from different Cartesian directions do not show large differences (see Appendix), therefore our conclusions are the same as Gray \cite{gray2022molecular}: PKA direction does not appear to strongly impact defect generation for collision cascades in YBCO.  \\

\subsection{Sub-Stoichiometric Collision Cascades}\label{substoichcascades}
\begin{figure*}
    \centering
    \includegraphics[width=\linewidth]{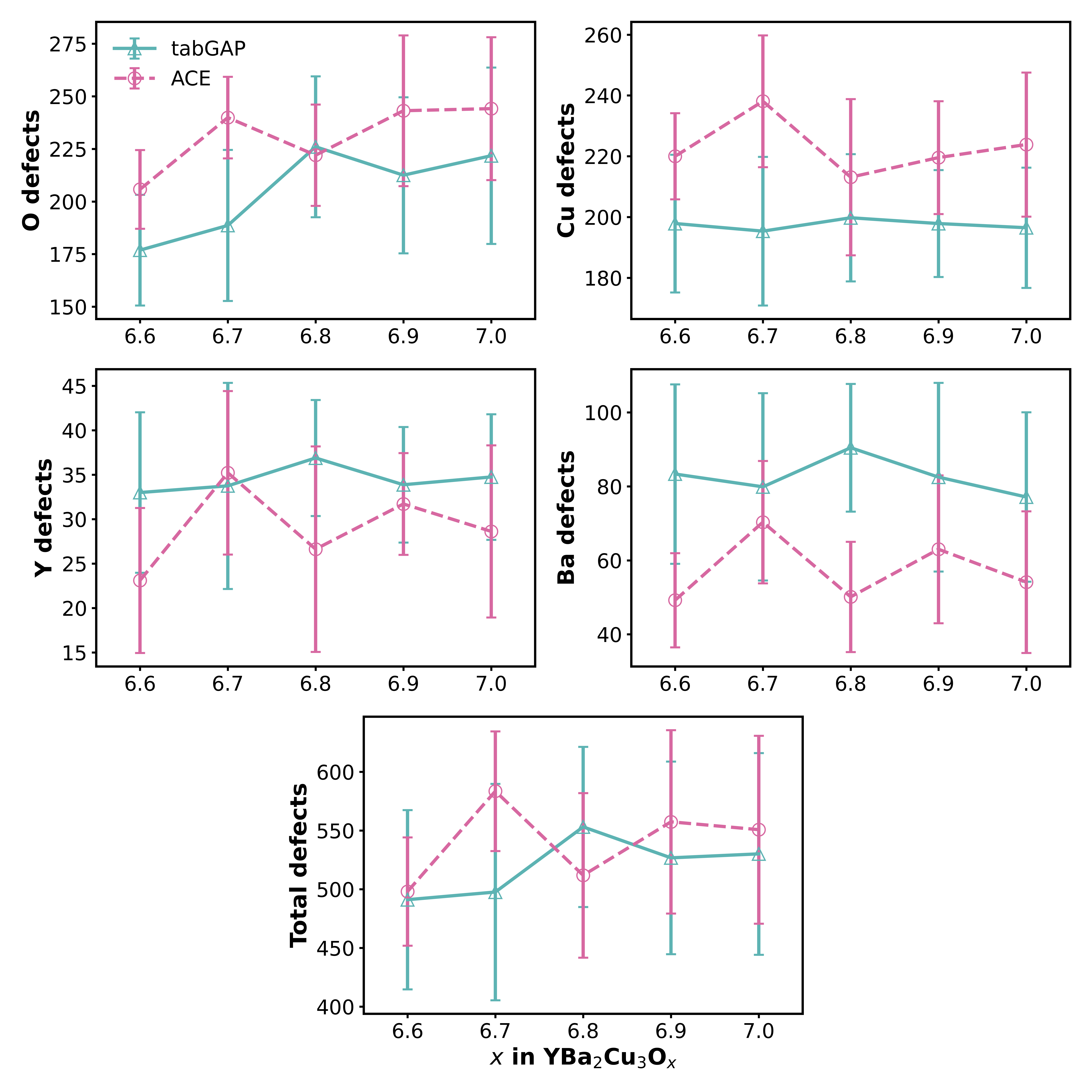}
    \caption{Defect count (antisites, interstitials, vacancies) against oxygen content in YBCO. The points are mean values for each stoichiometry and the error bars represent the standard deviation of the values. Data is shown from both ACE and tabGAP.}
    \label{defvsconc}
\end{figure*}

To demonstrate the ability of the MLPs to model sub-stoichiometric material, we have performed 5 keV Ba PKA cascade simulations three times in each Cartesian direction, YBa$_2$Cu$_3$O$_{7-\delta}$ for $0\leq\delta\leq0.4$. We sample this range so our results are comparable to experiments, as for YBa$_2$Cu$_3$O$_{6.6}$ the formation of more complex superlattice structures is favourable. For instance, the Ortho-II and Ortho-III phases, where chains are occupied with oxygen every two or three chains, respectively \cite{manca2000critical, manca2001orthorhombic}. Studying these phases is beyond the scope of the present work. We have tested the stability of the Ortho-II phase with the tabGAP and ACE and found that both models correctly predict this arrangement as more energetically favourable than randomly distributed oxygen in the chains. In order to calculate defect counts, the sub-stoichiometric supercell was compared to a perfect YBa$_2$Cu$_3$O$_{7}$ reference cell using Wigner-Seitz analysis. The initial defect count (before the cascade) then corresponds to the total number of oxygen defects in the undamaged sub-stoichiometric cell. Therefore, to obtain defect counts from the cascade, the initial defect count is simply subtracted from the calculated amount. Mean final defect counts for ACE and tabGAP are shown in Figure \ref{defvsconc}.\\

We argue that a defect count including interstitials and antisites (as well as vacancies) is more insightful for damage production. Therefore, for this section, antisite and interstitial defects are included in the total defect count. It is worth noting here that the detection of a defect (via Wigner-Seitz analysis) in an amorphous region cannot be interpreted as a point defect. The Wigner-Seitz algorithm will detect many interstitials (occupancy of greater than 1 for a given Voronoi cell), vacancies (Voronoi cell occupancy of 0), and antisites (Voronoi cell occupancy of 1, but the species occupying the site is different than the pristine cell). Therefore, our final defect count only partially reflects the number of point defects in the lattice. As such, it should be interpreted as a relative measure of damage, rather than an absolute value of point defects. To supplement this discussion, we have demonstrated the rarity of true antisite defects in the Appendix.\\

Overall the ACE and tabGAP models show similar results. It is clear from Figure \ref{defvsconc} that there is a reduction in oxygen defects with decreasing oxygen content. This is due to two different effects: (1) the probability of displacement of an oxygen atom decreases as oxygen deficiency increases, therefore fewer oxygen defects are observed. This is clearly not the only reason for an increase in oxygen defects with oxygen content as a linear fit to the data for O in Figure \ref{substoichcascades} shows an increase of 14 and 24 \% with the ACE and tabGAP, respectively. This does not line up with the 6 \% increase in oxygen concentration, therefore suggesting another effect: (2) examination of the peak and final oxygen defect counts reveals a small increase in oxygen recombination percentage with increasing oxygen deficiency (see Appendix). Note that recombination on longer time scales may be more significant than what is seen here. In any case these effects result in the overall defect count for both the ACE and tabGAP models decreasing with a decrease in oxygen content. There is no clear trend visible for the cation defect counts, which are similiar for all stoichiometries. It is worth noting that cascades are inherently chaotic and branching events. Therefore, the variability in results is considerable. While it appears that there are small jumps in defect counts in places, these are insignificant given the size of the error bars. Overall, consistent with the reduction in oxygen defects, the total number of defects appear to decrease with decreasing oxygen content. \\ 

\begin{figure}
    \centering
\includegraphics[width=1\linewidth]{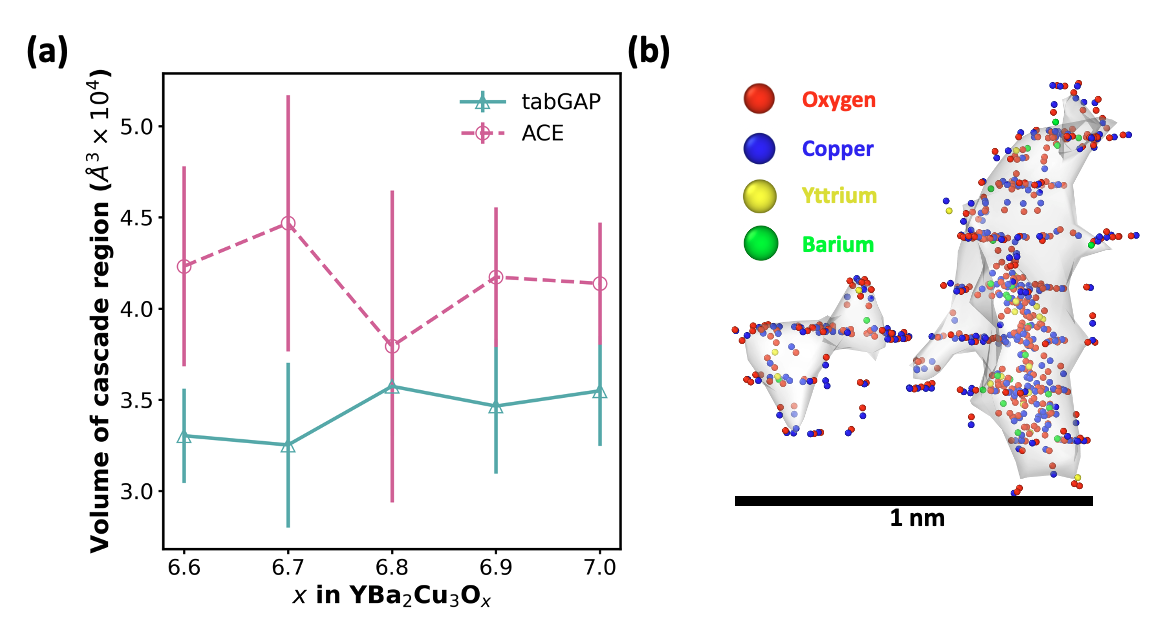}
    \caption{(a) Volume of damaged region (obtained by projecting a surface mesh onto defects resulting from cascades using Ovito \cite{stukowski2009visualization}) versus oxygen content in YBCO. The points are mean values for each stoichiometry and the error bars represent the standard deviation of the values. (b) Example of a volume projected onto identified defects from a 5 KeV cascade.}
    \label{vol}
\end{figure}

Figure \ref{vol} shows how the volume of the damaged region changes with the oxygen content in YBa$_2$Cu$_3$O$_{7-\delta}$. This provides an indication of how irradiation may affect the production of pinning centres. To estimate the volume of these regions, a surface mesh is projected onto the identified defects with the AlphaShape method in Ovito \cite{stukowski2009visualization}. This method is not deterministic as the results can change based on the settings used. We use the same parameters for all volume constructions and so it serves as a good indicator of how the volume changes. There is no clear trend in volume change as a function of oxygen content for the MLPs. The main difference is that the ACE appears to describe a larger damage volume, as well as a larger variability in the damage volume (this is likely tied to its higher standard deviation for defect counts). The larger volume predicted by ACE may be due to the inaccuracies it displayed in its initial fitting versus oxygen content. For lower oxygen contents, the ACE overestimates the $c$ axis expansion compared to the DFT and experimental standard, whereas the tabGAP follows the DFT tightly \cite{MLPYBCO}. Differences in the stiffness of the repulsive potentials (as shown in Figure \ref{QSD}b) may also give rise to these opposing results. To summarise, the results from both MLPs suggest that oxygen stoichiometry in YBCO tapes is unlikely to have a significant effect on the amount of damage observed, however, there may be a slight decrease in oxygen defect production with decreasing oxygen content. \\

\subsection{Fusion-Relevant Cascades and Simulated HRTEM Images}

\begin{figure}
    \centering
    \includegraphics[width=\linewidth]{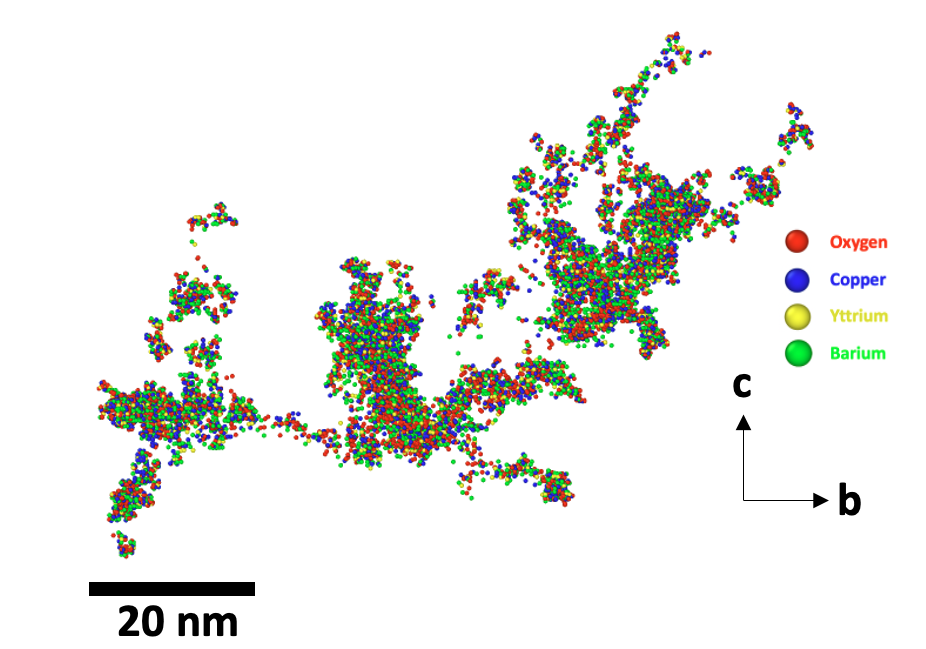}
    \caption{Snapshot (at time $\approx$ 300 fs) of heat spike in 300 keV Ba PKA cascade in $c$ direction of YBCO$_{6.93}$ at 25 K (using the tabGAP). Only atoms with kinetic energy greater than 0.5 eV are shown. We identified 27,949 defects (antisites, interstitials, vacancies) in the final frame of the cascade.}
    \label{cascade}
\end{figure}

In this final section, we perform a high energy, fusion-relevant collision cascade in YBa$_2$Cu$_3$O$_{6.93}$, which is a typical stoichiometry for commercial tapes. We have chosen a 300 keV Ba PKA as the highest energy Ba PKA we would expect to see from the ARC fusion reactor at the TF magnet coil \cite{torsello2025ieeetas}. For such a high energy event, the computational requirements are very demanding due the large supercell required (>350 million atoms). In light of its superior efficiency, the tabGAP is the most suitable potential for this task ($\sim5\times$ faster than ACE \cite{MLPYBCO}).
Moreover, given the high energy involved in this case, the approach distance between atoms will become small, meaning that the disagreement between the ACE short-range repulsion and DFT (as seen in Fig. \ref{QSD}b) may introduce inaccuracies.
A graphic representation of the cascade heat spike is shown in Figure \ref{cascade}.

A graphic representation of the cascade heat spike is shown in Figure \ref{cascade}. As evident from the figure there is significant cascade splitting, resulting in the formation of multiple sub-cascades, as is common at higher PKA energies \cite{sand2017cascade}. Note that we select the PKA direction as the $c$ axis. Given that most analyses of neutron irradiated samples come from nearly-isotropically irradiated tapes (for instance those from the TRIGA reactor \cite{fischer2018effect}), there is no clear choice for the PKA direction. In the toroidal field coils of a typical Tokamak style fusion reactor, the $c$ axis of the tape is orthogonal to the plasma surface, therefore, for application to fusion we chose the $c$ axis as the PKA direction. However, given that our prior results combined defect counts from PKAs directed along all Cartesian axes, and these results did not appear to change much with direction (consistent with the results from Gray \textit{et al.} \cite{gray2022molecular}), it is safe to assume the damage observed from different directions will not be hugely dissimilar. \\

\begin{figure}
    \centering
    \includegraphics[width=1\linewidth]{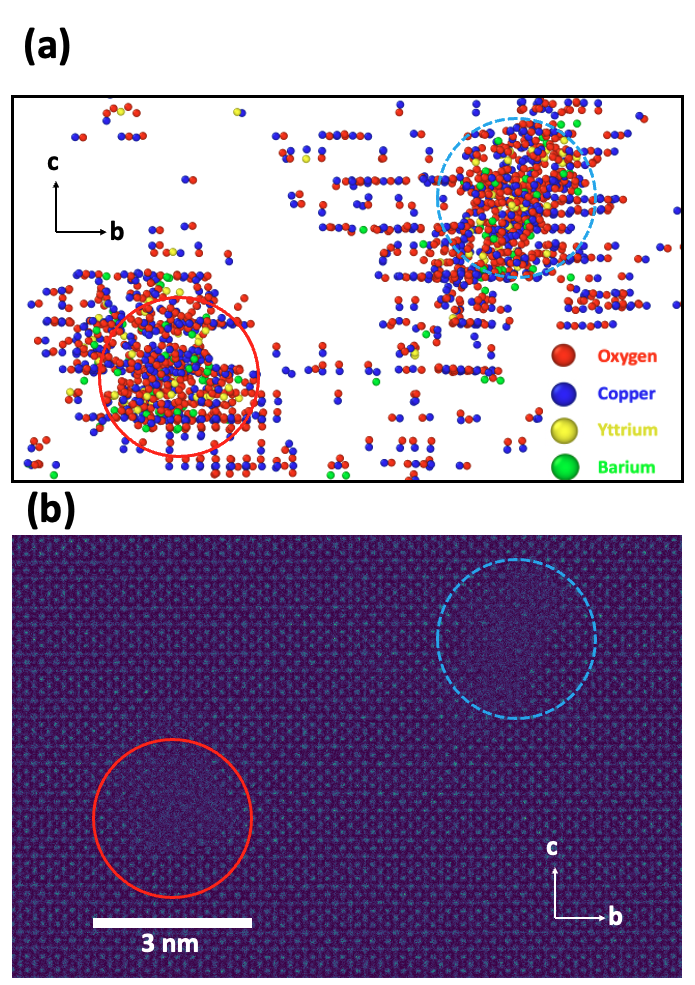}
    \caption{(a) Defects near amorphous cores from the conclusion of the cascade in figure \ref{cascade}. Two different amorphous cores are highlighted with a solid red circle and dashed blue circle.  (b) Simulated HRTEM image of a damage region in the 300 keV cascade from figure \ref{cascade}. Simulation performed using abTEM \cite{madsen2021abtem}. The amorphous regions are indicated with solid red and dashed blue circles, corresponding to the same amorphous cores in (a). These regions are approximately 3 nm across. The $bc$ plane is imaged, for direct comparison to \cite{linden2022analysing}. Note that both (a) and (b) are the same scale.}
    \label{abtem}
\end{figure}

Figure \ref{abtem}a shows the defects formed in amorphous regions from the above cascade. Similarly to what was observed using the Gray potential \cite{gray2022molecular}, the central regions of the cascade exhibit an amorphous structure, whereas the peripheral regions show defects predominantly in the CuO chains. These defects are mostly interstitial oxygen atoms located in the vacant $a$ axis position, orthogonal to the CuO chains. It is commonly thought that isolated point defects are the primary cause of superconductivity suppression under irradiation \cite{sauerzopf1998anisotropic, kirk1993structure, nicholls2022understanding, eisterer2025degradation}. Consistent with this, our simulations reveal regions of isolated point defects outside the peripheral zone of the amorphous core; however, their overall number is relatively low. Moreover, it remains uncertain whether these defects persist over longer timescales, as they may undergo annealing. Experimental evidence indicates that some of these point defects can indeed be annealed \cite{chudy2012point, Unterrainer_2022}.\\

A proposed simulation-driven workflow for prediction of radiation damage in HTS tapes from Torsello \textit{et al.} \cite{torsello2025roadmap} outlines the need for accurate prediction of the morphology of pinning centres from more accurate interatomic potentials. The results can then (in theory) be passed forward to Time-Dependent Ginzburg Landau theory \cite{milovsevic2010ginzburg, sadovskyy2016simulation, sadovskyy}, to predict the resulting effect on superconducting performance. These MLPs were created to fill this need. Therefore, we have constructed a simulated HRTEM image of the damaged region from the above cascade to determine the accuracy of our results compared to experimental images of neutron irradiated tapes from Linden \textit{et al.} \cite{linden2022analysing}. Note that these irradiations were performed in the TRIGA reactor in Wien, whose spectrum has a fast neutron component that covers part of the spectrum generally expected in compact fusion reactors \cite{torsello2025ieeetas}. Our simulated HRTEM image is shown in Figure \ref{abtem}b. \\

The simulated HRTEM image displays roughly spherical amorphous regions on the order of $\xi$ (a few nm), which is shown to exhibit effective pinning \cite{zhang2022progress}. This is therefore consistent with the observation of an initial increase in the tape critical current ($I_c$) under neutron irradiation (demonstrated by the TU Wien group \cite{fischer2018effect, Unterrainer_2024}). Furthermore, TEM images of neutron irradiated YBCO single crystals exhibit amorphous regions with an average size of 3 nm \cite{frischherz1994defect}, in direct agreement with our cascade. We do however note that the size of the amorphous regions exhibited in HRTEM images of neutron irradiated GdBCO from Linden \textit{et al.} \cite{linden2022analysing} are around 10 nm. We studied (as a proof of concept) a single 300 keV Ba-initiated cascade, whereas in a TRIGA experiment the damage comes from a broad distribution of PKAs, potentially explaining the observed discrepancy. Given the large spread in TEM data, controlled experimental studies in comparison with simulations are required to glean more information on the morphology of damage cascades. In any case, the size of amorphous regions here are in excellent agreement with the currently available experimental observations. \\

\section{Conclusion}

We have assessed two recently developed machine learned potentials (ACE and tabGAP) on their ability to model radiation damage in YBCO across both stoichiometric and oxygen-deficient compositions. We show that both models achieve near-DFT accuracy for the key physical processes underpinning displacement cascades. Quasi-static drag calculations confirm that the curvature of the potential-energy surface is well described by both MLPs, with the tabGAP showing slightly more accurate results for the higher energy regime. Both MLPs represent a significant improvement over the Gray potential~\cite{gray2022molecular}. Threshold displacement energies of O1 for both ACE and tabGAP agree quantitatively with DFT predictions. Liquid-phase RDFs demonstrate that both MLPs correctly reproduce the liquid phase, important for the accurate modelling of the heat-spike of a collision cascade. \\

Cascade simulations show the same overall structure as previous simulations using the Gray potential, amorphous regions surrounded with  defects in the CuO chains. However, the MLPs consistently generate higher peak defect counts than the Gray potential but also substantially greater recombination. Furthermore, the final defect counts are larger for the MLPs than the Gray potential, and the production of copper defects is increased, agreeing with experimental observation of Cu-O divacancies. ACE and tabGAP give similar average counts, however, the standard deviation of defect counts for ACE is larger than that of tabGAP (perhaps due to issues with the short range repulsion). We examined defect numbers as a function of PKA direction and found no strong directional dependence. The MLPs also successfully describe sub-stoichiometric YBCO, enabling the first systematic computational investigation of radiation damage as a function of oxygen deficiency. We find that oxygen depletion slightly reduces oxygen defect production and may modestly alter the cascade-damaged volume, but overall defect levels remain relatively insensitive to stoichiometry in the YBCO$_{6.6}$–YBCO$_7$ range.\\

Finally, fusion relevant 300~keV cascade simulations (with the tabGAP) and HRTEM emulation reveal amorphous regions a few nanometres in diameter, consistent with experimentally observed pinning centres and comparable to the coherence length. This demonstrates that the tabGAP reproduces the scale and morphology of irradiation-induced defects relevant to superconducting performance. Together, these results establish the ACE and tabGAP models as robust and efficient tools for predictive radiation-damage modelling in YBCO; although for higher energy cascades we recommend the tabGAP due to its improved short range interactions. Furthermore, the computational speed of the tabGAP (5$\times$ faster than the ACE) allows for larger scale simulations. However, given the more accurate defect properties of the ACE, longer timescale defect annealing may be less accurate with the tabGAP. Both the ACE and tabGAP MLPs represent a significant improvement upon all prior potentials for YBCO, underscoring the power of MLPs for modelling complex radiation damage processes.

\section*{Acknowledgements} 

DFT calculations have been performed by AD on the CSD3 HPC. Calculations with the potentials were performed by AD on CSD3 and Lancaster University's HEC. Tuning of the ACE core repulsion potential was performed by NDE on Eni's HPC6.

AD would like to acknowledge that this project was funded by Lancaster University and UKAEA via agreement 2077239. DNM and MRG acknowledge funding from the EPSRC Energy Programme [grant number EP/W006839/1]. NDE acknowledges that this publication is part of the project PNRR-NGEU which has received funding from the MUR – DM 117/2023. NDE, DT, FeL and FL acknowledge support from Eni S.p.A. DG acknowledges financial support from the Swedish Research Council (VR) through Grant No. 2023-00208. This work was carried out (partially) using supercomputer resources provided under the EU-JA Broader Approach collaboration in the Computational Simulation Centre of International Fusion Energy Research Centre (IFERC-CSC)

\bibliographystyle{elsarticle-num} 
\bibliography{lib.bib}

\appendix
\section{Short range repulsion for GAP and tabGAP}\label{sec:dmol}

For the purposes of high energy collision cascades, we fit an all electron DFT informed short range pair potential to spline to the GAP \cite{nordlund2025repulsive}. This new type of short range potential is shown to better reproduce experimental ion implantation depths than the standard ZBL. It is presented in the form of a refitted ZBL potential:

\begin{equation}
    V_{pairs}(r_{ij}) = \frac{1}{4\pi \epsilon_0}\frac{Z_iZ_je^2}{r_{ij}}\phi(r_{ij}/a)f^{pairs}_{cut}(r_{ij})
\end{equation}
where $a$ takes the form:
\begin{equation}
    a= \frac{0.46848}{Z_i^{0.23} + Z_j^{0.23}}
\end{equation}

We utilise a cutoff function of the form shown in Eq. \ref{eqna3}, as suggested in Byggmastar \emph{et al}.\cite{byggmastar2019machine}:
\begin{equation}\label{eqna3}
   f^{\text{pairs}}_{\text{cut}}(r_{ij}) = 
\begin{cases}
1, & r_{ij} \leq r_1 \\
1 - \chi^3 (6\chi^2 - 15\chi + 10), & r_1 \leq r_{ij} \leq r_2 \\
0, & r_{ij} \geq r_2
\end{cases}
\end{equation}
for $\chi = (r_{ij}-r_1)/(r_2-r_1)$. The different cutoff parameters ($r_1$ and $r_2$) are chosen to provide a good fit to the all electron DMol data, and also must vanish well below the equilibrium bond distance, leaving the rest of the interaction energy to be machine learned. For each element pair we fit a screening function ($\phi$), which takes the form of a double exponential:
\begin{equation}
    \phi(x) = A_1 \text{exp}(-B_1x) + (1-A_1)\text{exp} (-B_2x)
\end{equation}

The free parameters are fitted using SciPy curve fit. The fitted parameters are shown in Table \ref{tab:pairs}.\\
\begin{table}[h]
\centering
\begin{tabular}{lccccc}
\hline
\textbf{Pair} & \textbf{A\textsubscript{1}} & \textbf{B\textsubscript{1}} & \textbf{B\textsubscript{2}} & \textbf{r\textsubscript{1}} & \textbf{r\textsubscript{2}} \\
\hline
O-O     & 0.1932313  & 2.9075653  & 0.6358699  & 0.5   & 1    \\
Ba-Ba   & 0.8464003  & 0.8631403  & 0.2651982  & 1.8   & 2.75 \\
Y-Y     & 0.689483   & 1.1980069  & 0.3458894  & 1.5   & 2.25 \\
Cu-Cu   & 0.6544874  & 1.1295844  & 0.4050281  & 1     & 2    \\
Cu-Ba   & 0.6502327  & 1.2237666  & 0.3823601  & 1.5   & 2.25 \\
Cu-Y    & 0.6109906  & 1.2941402  & 0.4079618  & 1.5   & 2    \\
O-Ba    & 0.7806831  & 0.9895612  & 0.3138540  & 0.75  & 1.5  \\
O-Cu    & 0.8546443  & 0.9146256  & 0.2991823  & 0.5   & 1.5  \\
O-Y     & 0.8270992  & 0.9514228  & 0.2727367  & 0.5   & 1.5  \\
Y-Ba    & 0.8154615  & 0.9515396  & 0.2857777  & 1.5   & 2.5  \\
\hline
\end{tabular}
\caption{Pair interaction parameters for various element combinations.}
\label{tab:pairs}
\end{table}

\begin{figure*}
    \centering
    \begin{subfigure}{0.32\textwidth}
        \includegraphics[width=\linewidth]{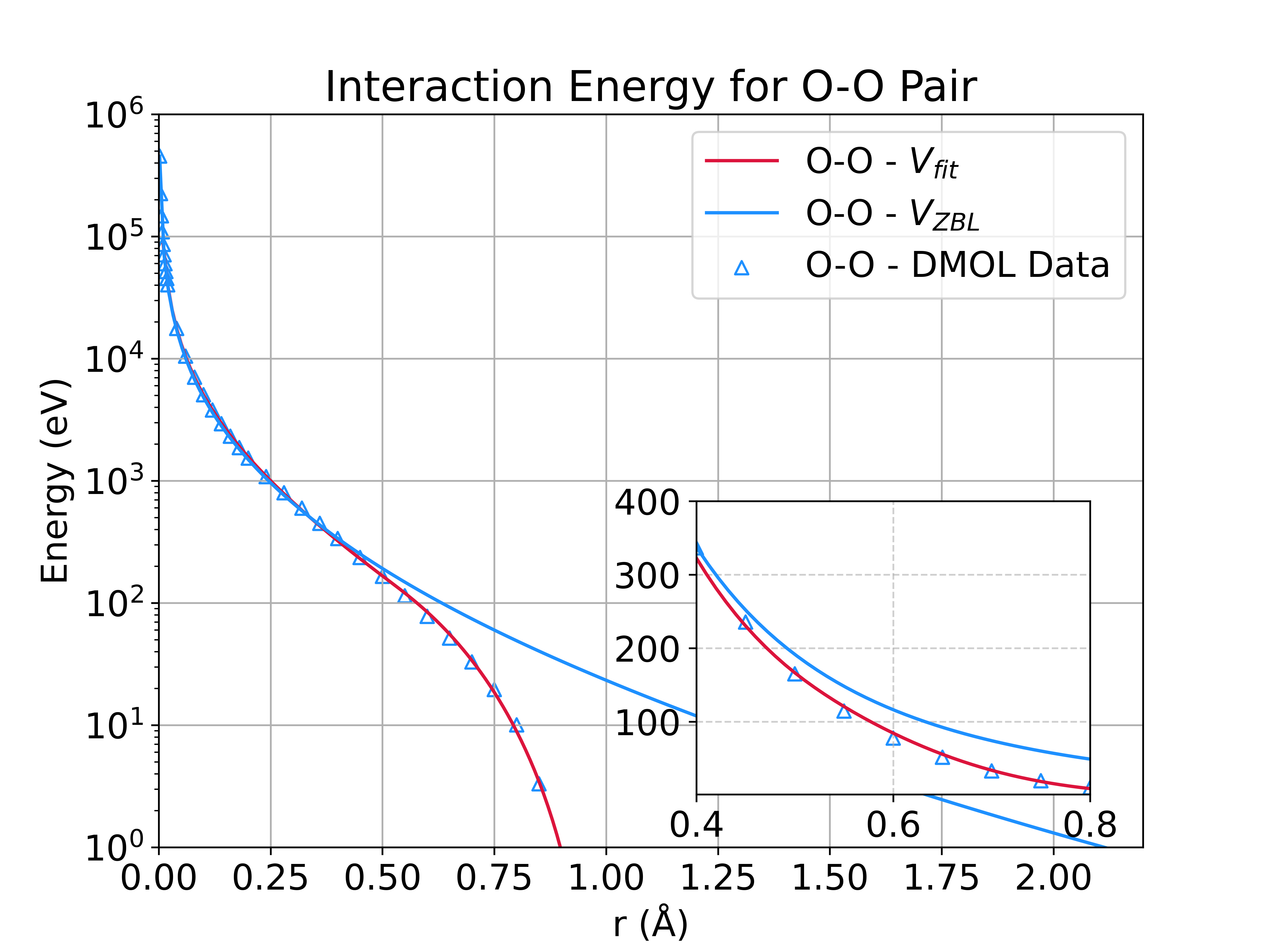}
        \caption{O–O}
    \end{subfigure}
    \begin{subfigure}{0.32\textwidth}
        \includegraphics[width=\linewidth]{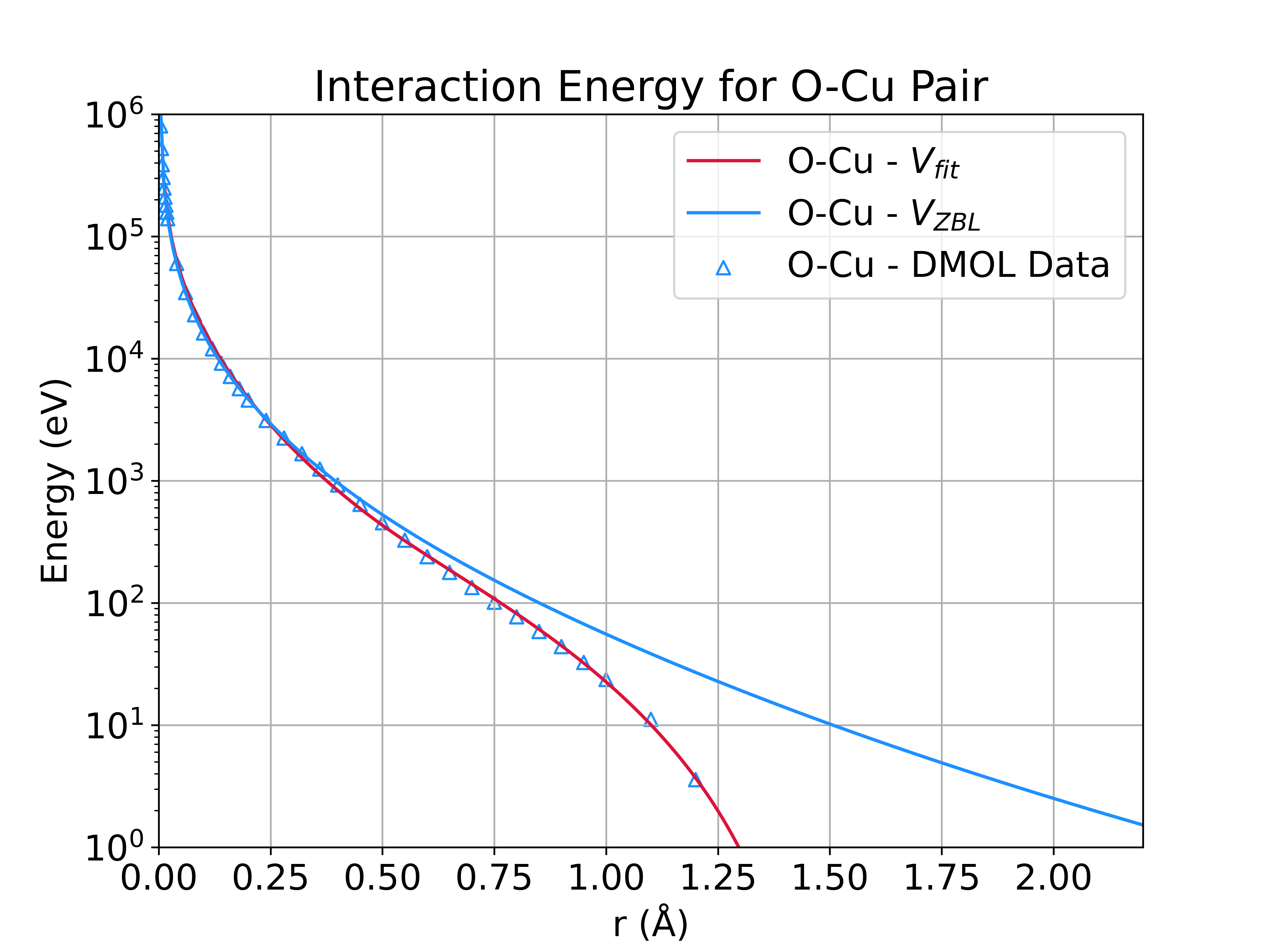}
        \caption{O–Cu}
    \end{subfigure}
    \begin{subfigure}{0.32\textwidth}
        \includegraphics[width=\linewidth]{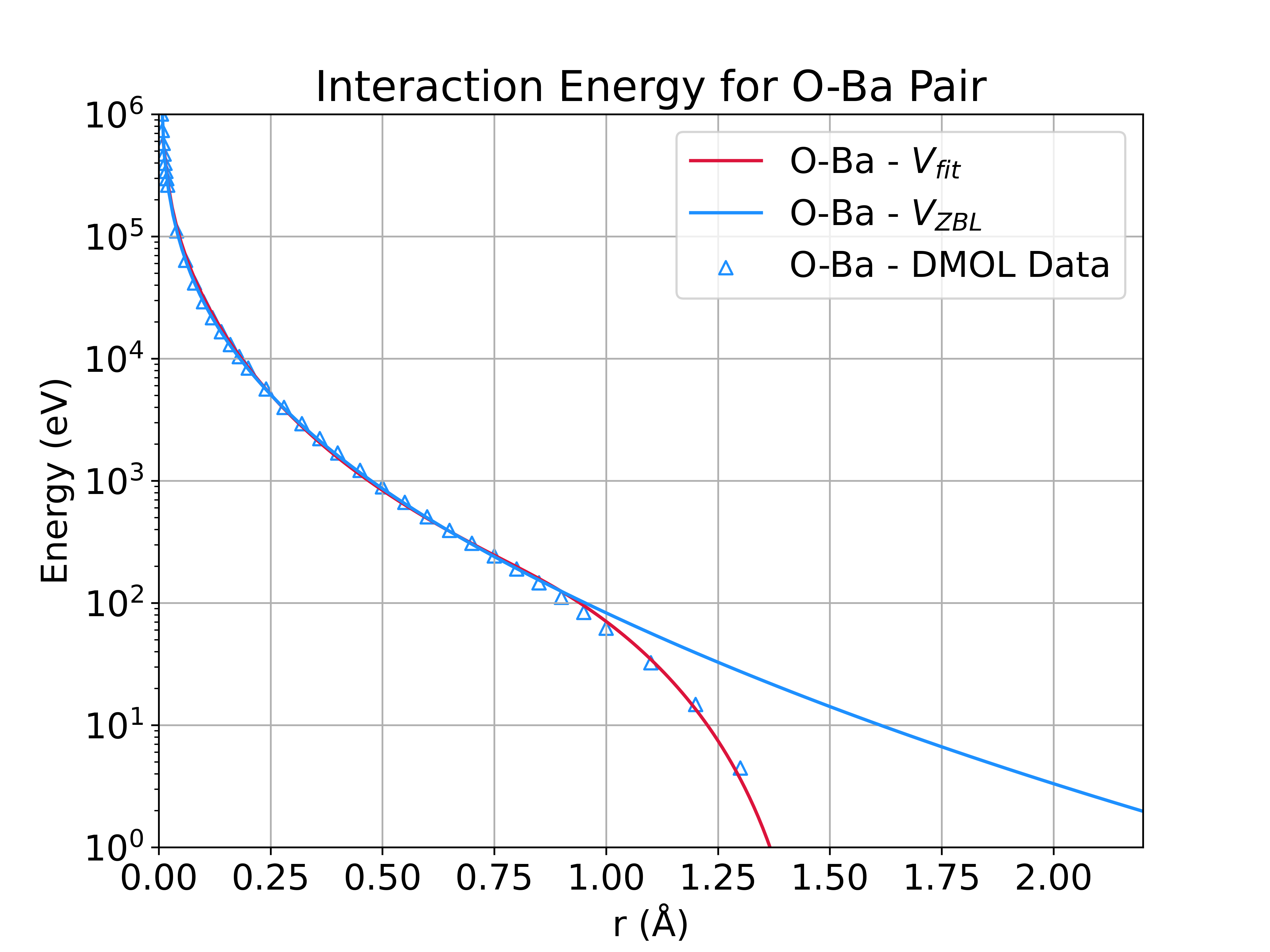}
        \caption{O–Ba}
    \end{subfigure}

    \begin{subfigure}{0.32\textwidth}
        \includegraphics[width=\linewidth]{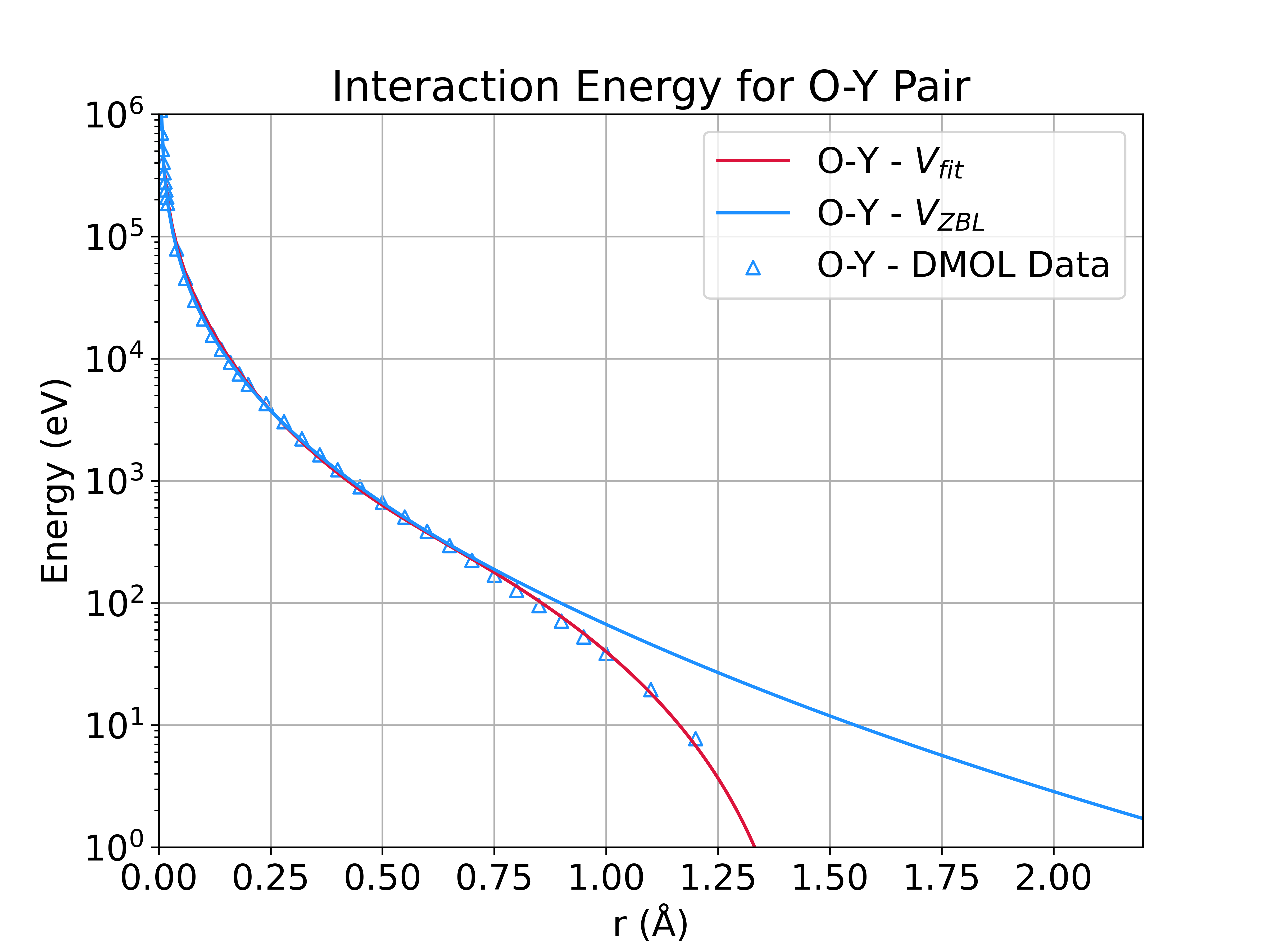}
        \caption{O–Y}
    \end{subfigure}
    \begin{subfigure}{0.32\textwidth}
        \includegraphics[width=\linewidth]{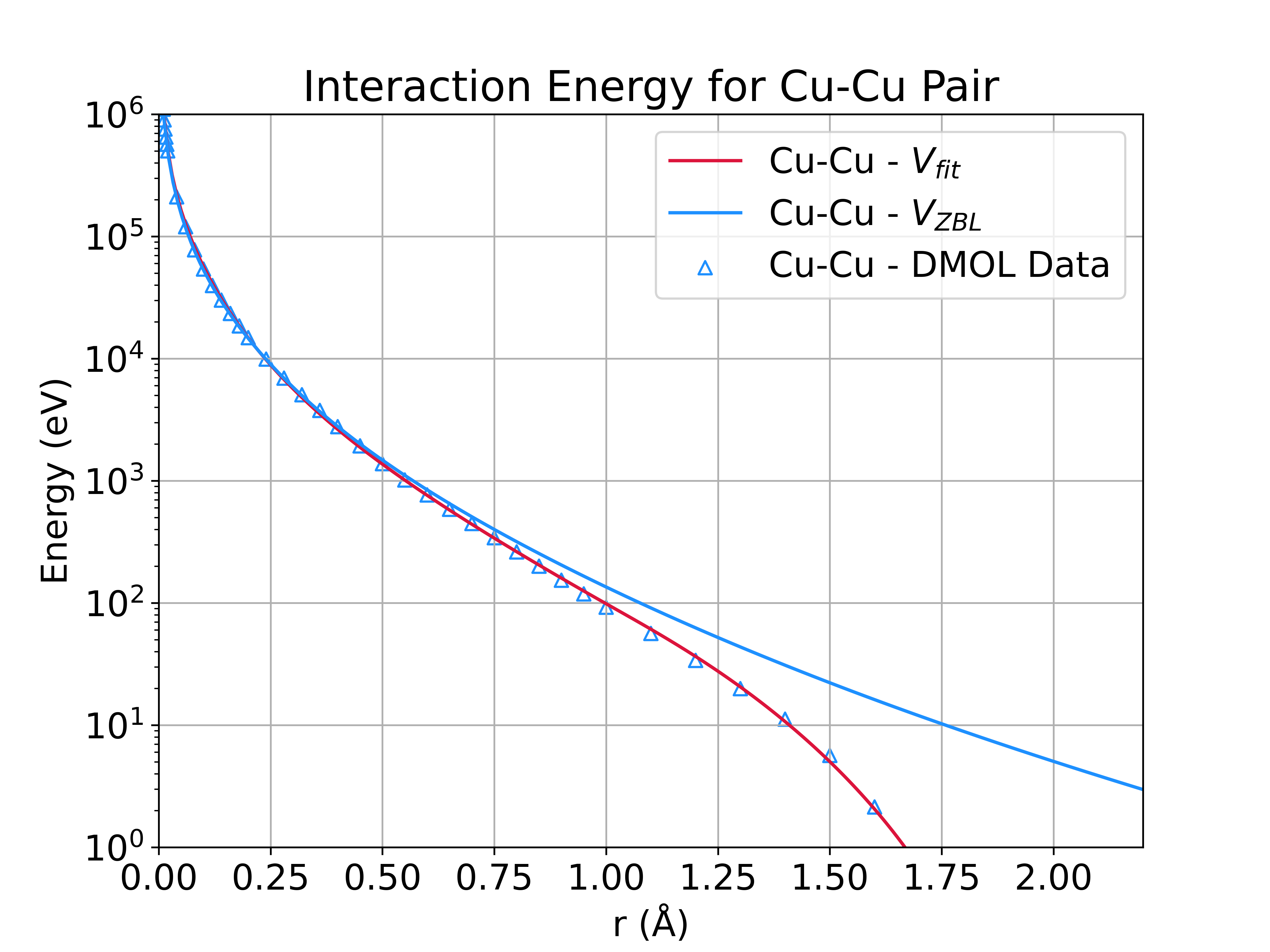}
        \caption{Cu–Cu}
    \end{subfigure}
    \begin{subfigure}{0.32\textwidth}
        \includegraphics[width=\linewidth]{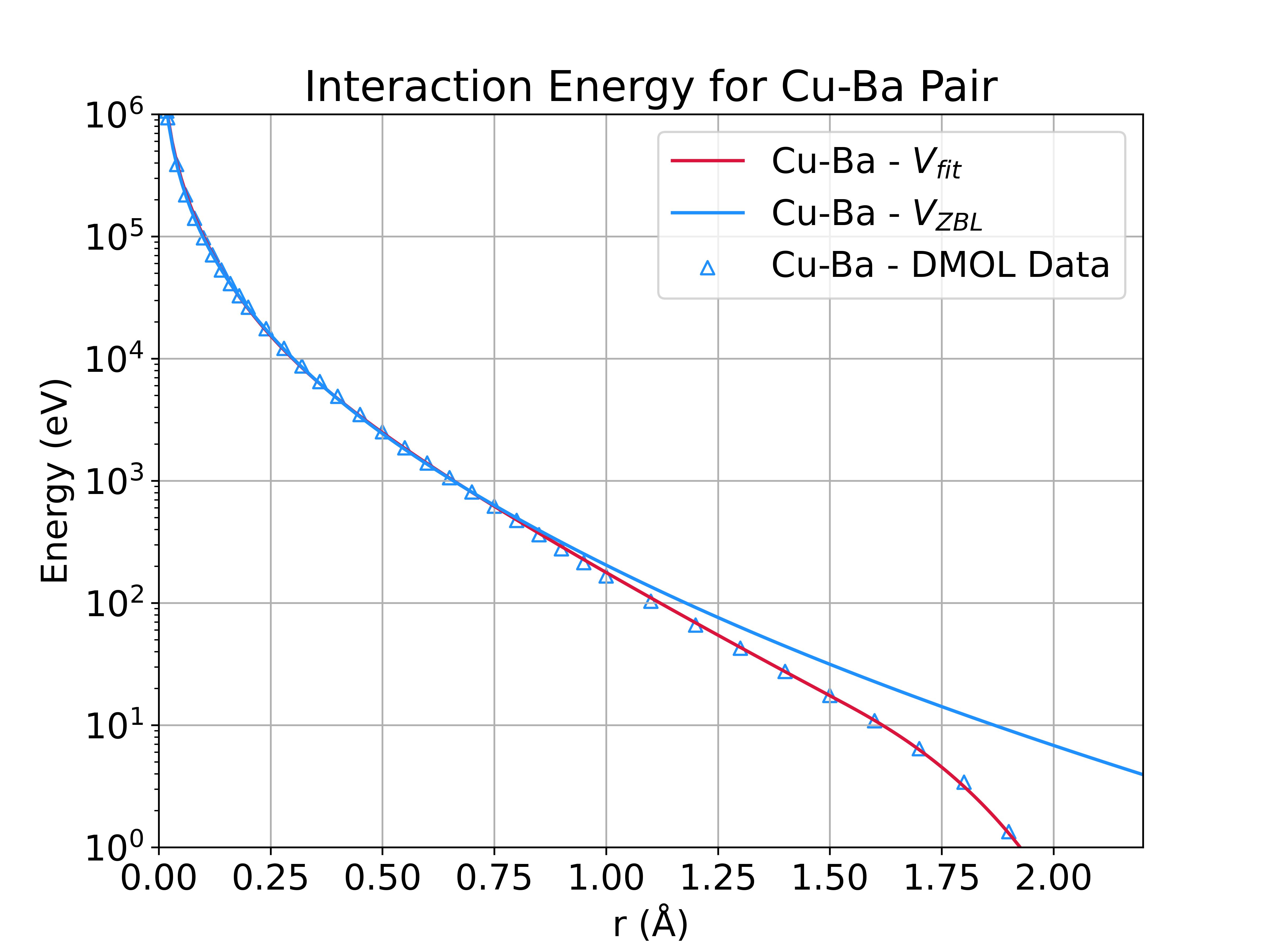}
        \caption{Cu–Ba}
    \end{subfigure}

    \begin{subfigure}{0.32\textwidth}
        \includegraphics[width=\linewidth]{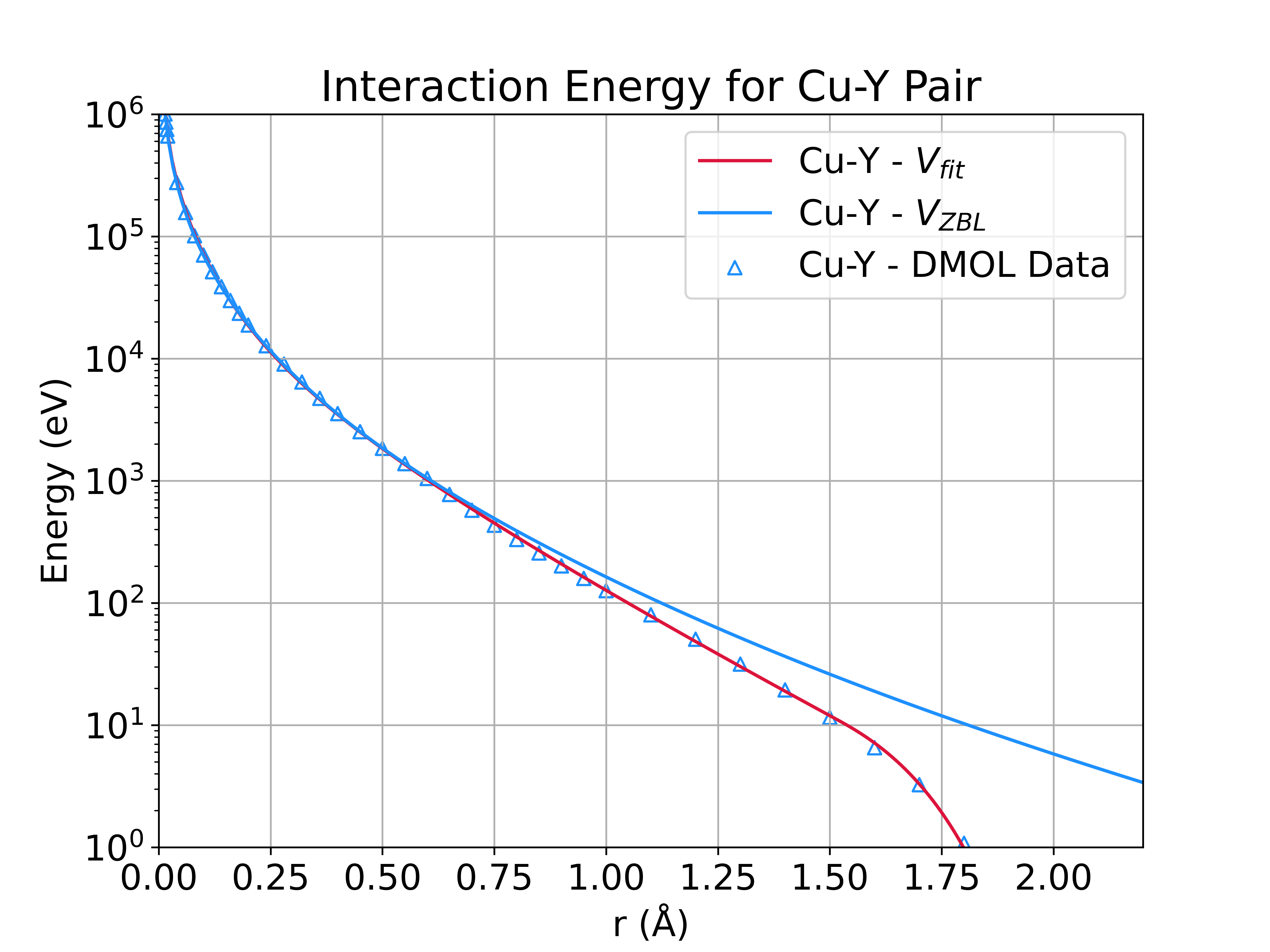}
        \caption{Cu–Y}
    \end{subfigure}
    \begin{subfigure}{0.32\textwidth}
        \includegraphics[width=\linewidth]{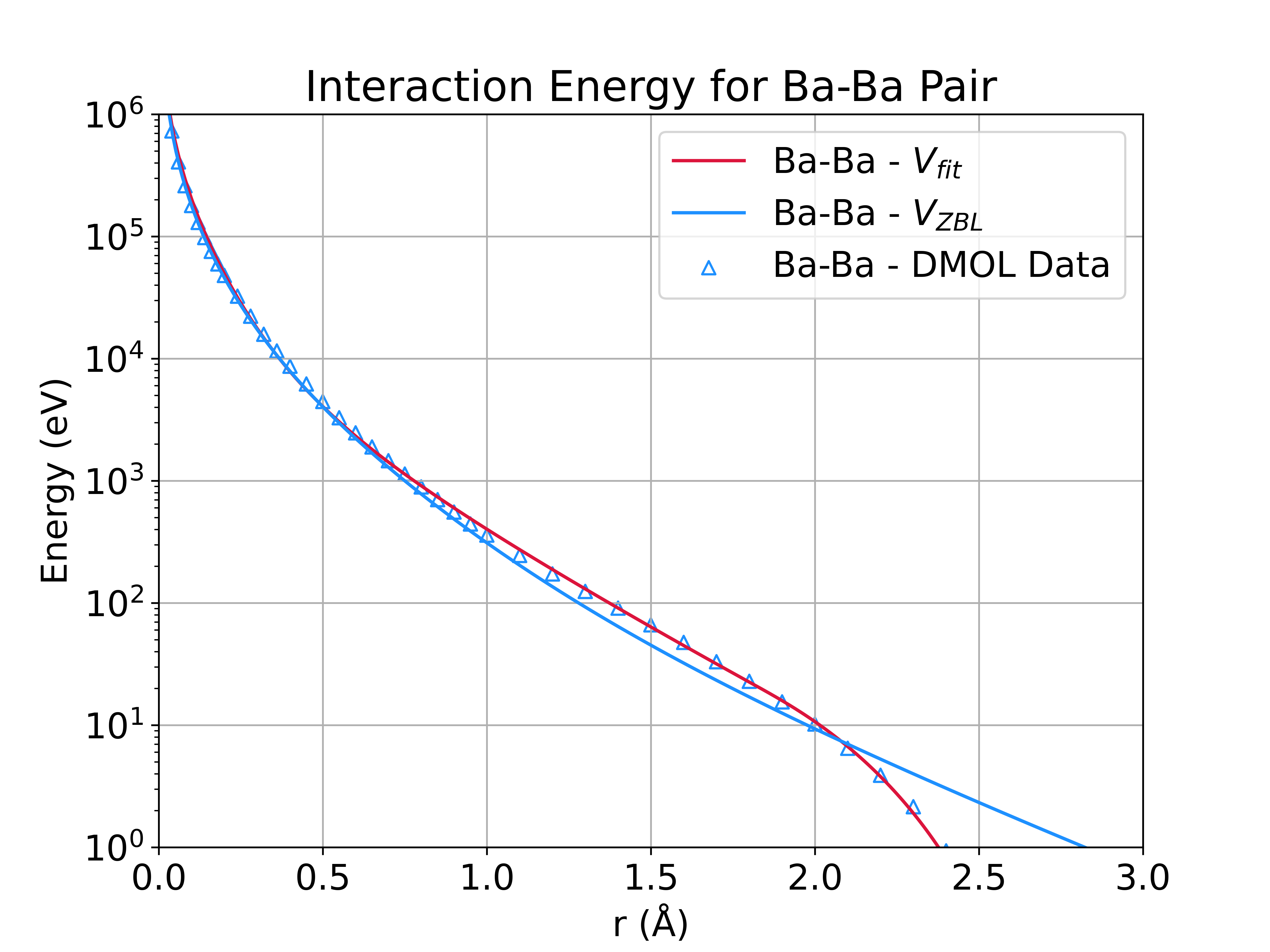}
        \caption{Ba–Ba}
    \end{subfigure}
    \begin{subfigure}{0.32\textwidth}
        \includegraphics[width=\linewidth]{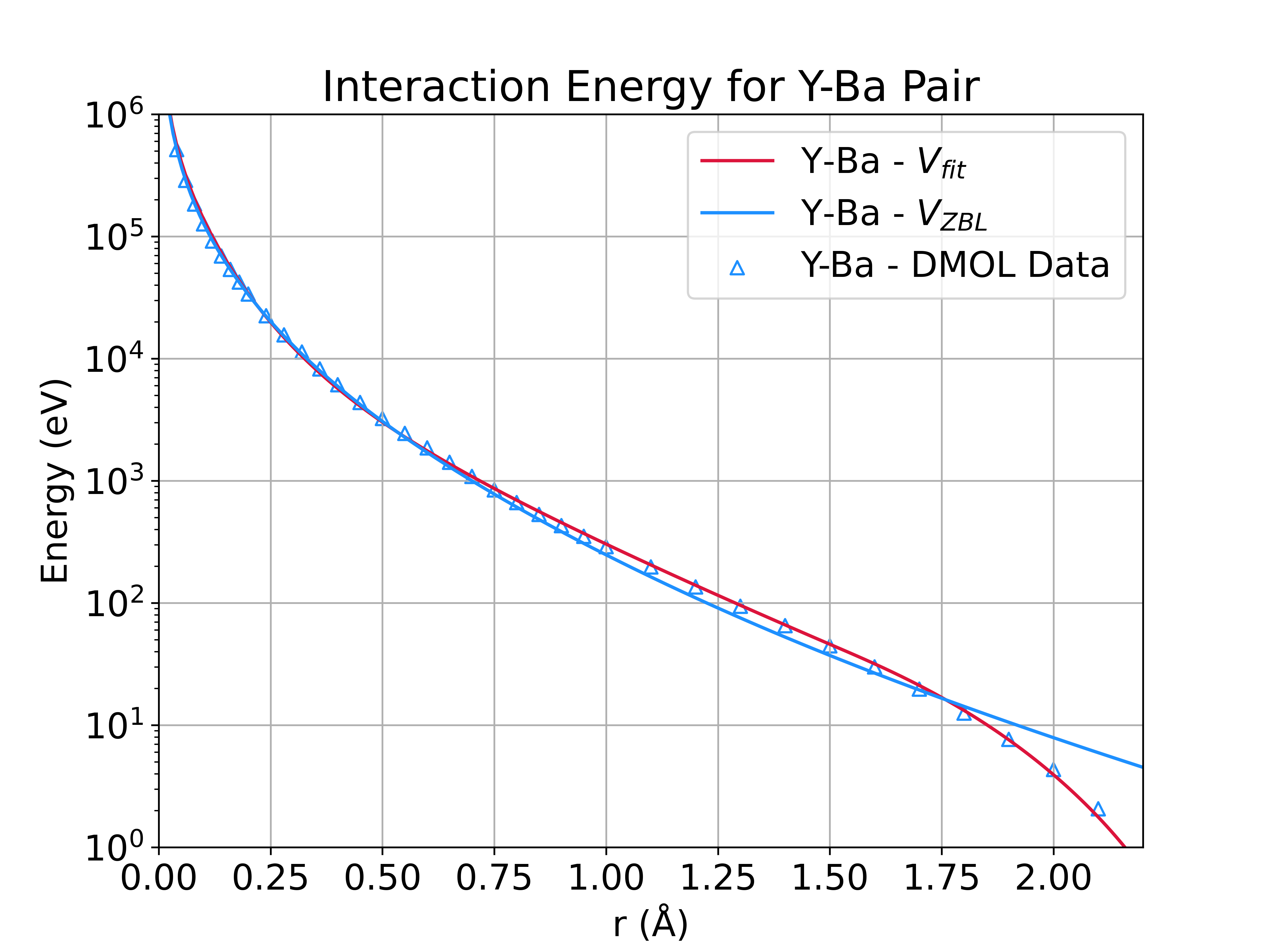}
        \caption{Ba–Y}
    \end{subfigure}

    \begin{subfigure}{0.32\textwidth}
        \includegraphics[width=\linewidth]{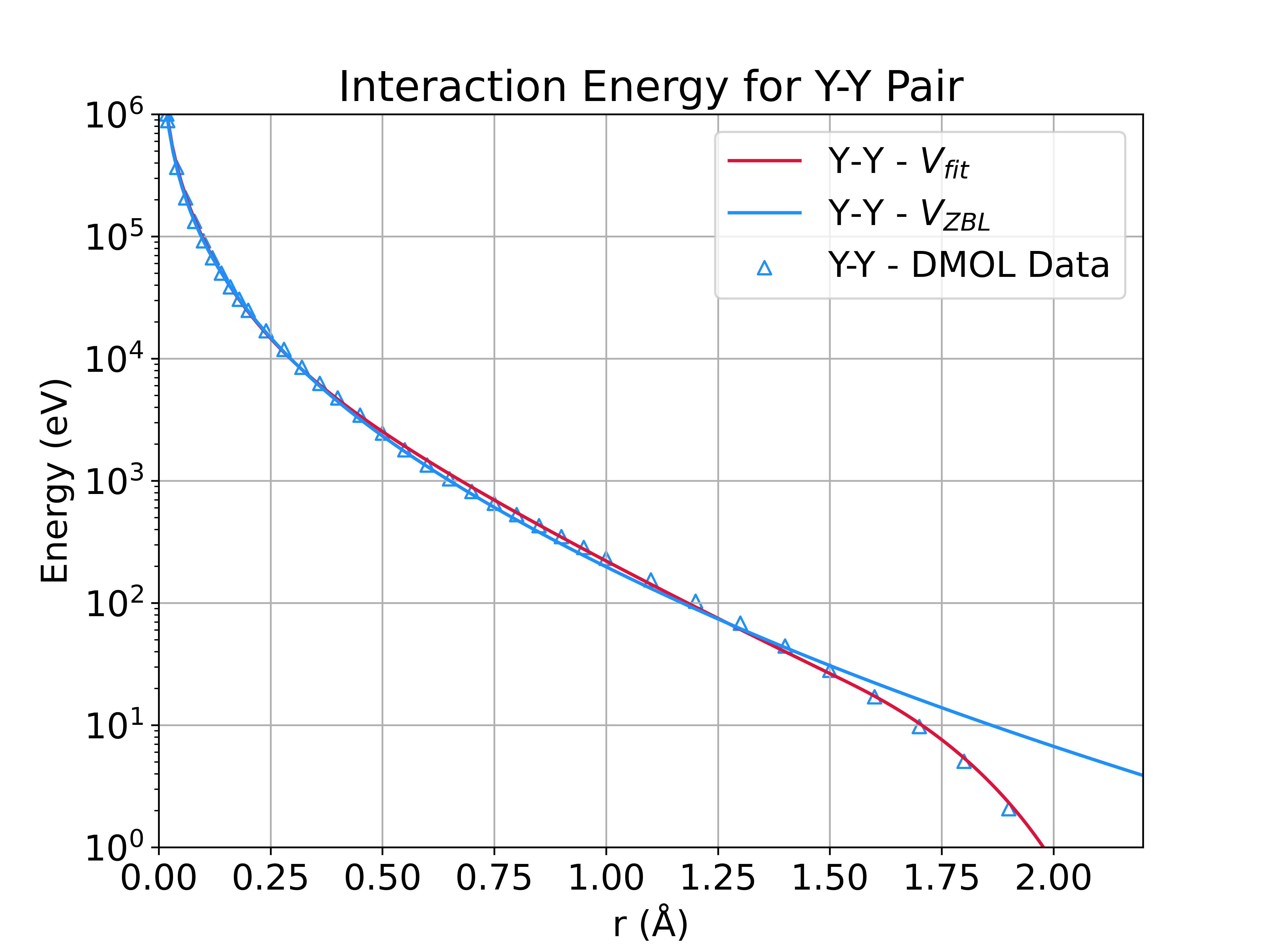}
        \caption{Y–Y}
    \end{subfigure}

    \caption{Fitted short range pair potentials for all unique element pairs. The solid red line is the fitted screened coulomb potential with the cutoff function included. The standard ZBL is also shown for comparison. }
    \label{dmolpair}
\end{figure*}

To demonstrate the difference between the short range pair potential and the ZBL, we have plotted the different functions in figure \ref{dmolpair}. In most cases, the ZBL and DMol fit show different behaviour for the 10-100 eV range. For higher energies the difference is harder to discern given the magnitude of the energies involved. The dimer curves for each pair are shown in figure \ref{fig:ace_tabgap_pairs}, demonstrating that the potential energy surface remains smooth and continuous after joining the GAP to the short range term.

\section{ACE Core Repulsion Potential}

The ACE potential used in this work was originally trained without an explicitly 
defined short-range repulsive term. Although the training input specified 
\texttt{repulsion:auto}, this option only restricts the domain of the spline basis and 
does not enforce physically meaningful behaviour at compressed interatomic 
separations. Because the training database contains no configurations probing these 
distances, the resulting model necessarily extrapolates in this regime.

To obtain a more controlled behaviour at short range, we supplemented the fitted ACE 
model with an analytic repulsive term introduced \emph{after} training. The functional 
form of this term follows that used in the tabGAP short-range correction described in 
~\ref{sec:dmol}, in which a screened Coulomb form is smoothly switched off between 
two radii $r_1(i,j)$ and $r_2(i,j)$. These radii were mapped to the ACE inner-cutoff 
parameters according to
\begin{equation}
r_{\mathrm{in}}(i,j) = r_1(i,j),
\end{equation}
\begin{equation}
\Delta r(i,j) = r_2(i,j) - r_1(i,j),
\end{equation}
where $\Delta r(i,j)$ corresponds to the ACE field \texttt{delta\_in}.  
The analytic prefactor and decay rate associated with the short-range term for each 
pair, denoted $P(i,j)$ and $\Lambda(i,j)$, were inserted into the ACE field 
\texttt{core\_rep\_parameters}:
\begin{equation}
\texttt{core\_rep\_parameters}(i,j)
= \bigl[\, P(i,j),\, \Lambda(i,j) \,\bigr].
\end{equation}
For consistency with the switching radii, the density-cutoff parameters were set equal 
to $r_2(i,j)$,
\begin{equation}
\rho_{\mathrm{cut}}(i,j) = r_2(i,j),
\qquad
\Delta\rho_{\mathrm{cut}}(i,j) = r_2(i,j).
\end{equation}

After these post-training modifications, the effective pair interaction becomes \begin{equation} 
\begin{aligned} V_{\mathrm{ACE\text{-}SR}}(i,j)(r) = \begin{cases} V_{SR}(i,j)(r), & \text{if } r \le r_1(i,j), \\[8pt] \begin{aligned} &\left[1 - S(i,j)(r)\right]\, V_{SR}(i,j)(r) \\ &\quad {}+ S(i,j)(r)\, V_{ACE}(i,j)(r), \end{aligned} & \text{if } r_1(i,j) < r < r_2(i,j), \\[12pt] V_{ACE}(i,j)(r), & \text{if } r \ge r_2(i,j). \end{cases} \end{aligned} 
\end{equation} 
where $V_{SR}(i,j)(r)$ is the DMol-derived short-range potential of \ref{sec:dmol}, and $S(i,j)(r)$ is the switching function acting on the interval $[r_1(i,j),r_2(i,j),]$.

While the construction above follows the tabGAP short-range formalism, we emphasise that embedding the same parameters into ACE does not reproduce the original DMol dimer curves exactly. This mismatch is expected: in tabGAP the short-range potential entirely replaces the extrapolative behaviour of the GAP, whereas in ACE the many-body structure of the basis and its extrapolation outside the training domain interact with the inserted repulsion. Consequently, the short-range behaviour reflects a combination of the analytic term and the ACE extrapolation, rather than a strict reproduction of the DMol reference data.

\section{Parity Plots}

\begin{figure*}[t]
    \centering
    \includegraphics[width=0.95\textwidth]{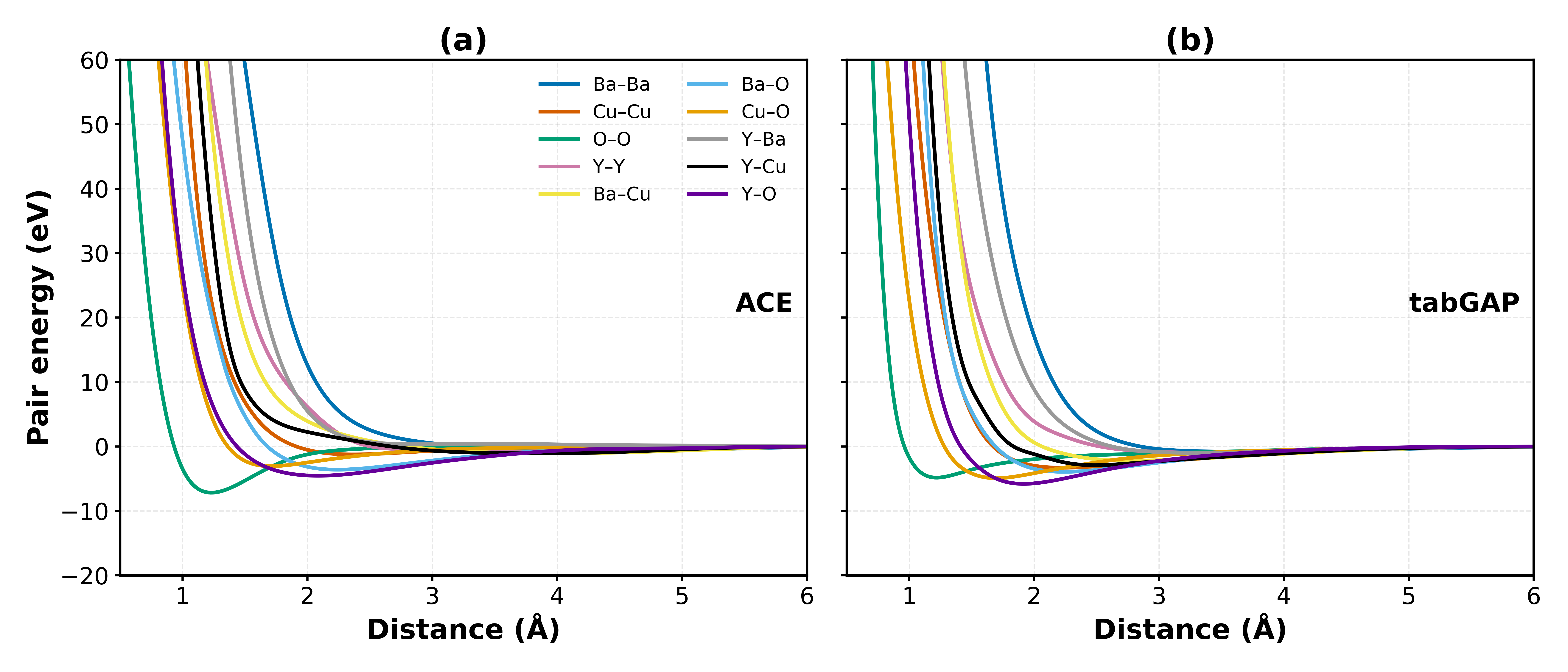}
    \caption{
    Short–range pair potentials for all unique element pairs in 
    YBa$_2$Cu$_3$O$_{7-\delta}$.  
    Panel~(a) shows the pair interactions obtained with the ACE potential, 
    while panel~(b) shows the corresponding DMol–fitted tabGAP.
    }
    \label{fig:ace_tabgap_pairs}
\end{figure*}

To analyse the short–range behaviour of both the ACE and tabGAP potentials, we computed
pair–interaction curves for all unique element pairs in 
YBa$_2$Cu$_3$O$_{7-\delta}$.
The procedure was identical for ACE and tabGAP, differing only in the underlying
calculator used to evaluate the energies. The results are shown in Fig. \ref{fig:ace_tabgap_pairs}.

For each chemical species  ${\rm X} \in \{{\rm Y, Ba, Cu, O}\}$,  the isolated–atom energy $E^{\rm iso}_{\rm X}$ was first computed by evaluating a  single–atom configuration using the corresponding potential.  
These values provide the reference required to extract meaningful two–body energies.
Then, for each pair $(i,j)$, a linear two–atom dimer
\[
{\rm X}_i{\rm X}_j(r) = \bigl\{ (0,0,0),\, (0,0,r) \bigr\}
\]
was constructed and evaluated on a uniform radial grid
\[
r = 0.10,\; 0.11,\; \ldots,\; 9.99\ \text{\AA},
\]
corresponding to a spacing of 0.01\,\AA.  
At each distance, the two–body interaction was obtained via
\[
E_{\rm pair}(i,j;r)
=
E_{\rm dimer}(i,j;r)
\;-\;
E^{\rm iso}_i
\;-\;
E^{\rm iso}_j ,
\]

For plotting, the raw dimer data were interpolated using a cubic B--spline for visual smoothness; but no smoothing or additional fitting was applied to the  underlying potentials.  
Both ACE and tabGAP curves were displayed over the interval $0.5$--$6$\,\AA\ which is sufficient to show that the pair energies return to zero.

\section{Antisite Defects in Cascades}

As discussed in the main text, we have included antisite defects in the total defect count to obtain a more realistic estimate of damage. However, we do not distinguish between the types of antisite defect, as most arise due to amorphisation in the material. This can be concluded by folding a simulation supercell into the unit cell after a collision cascade. This has been performed for one cascade and is shown in Supplementary figure \ref{folded}.

\begin{figure}
    \centering
    \includegraphics[width=\linewidth]{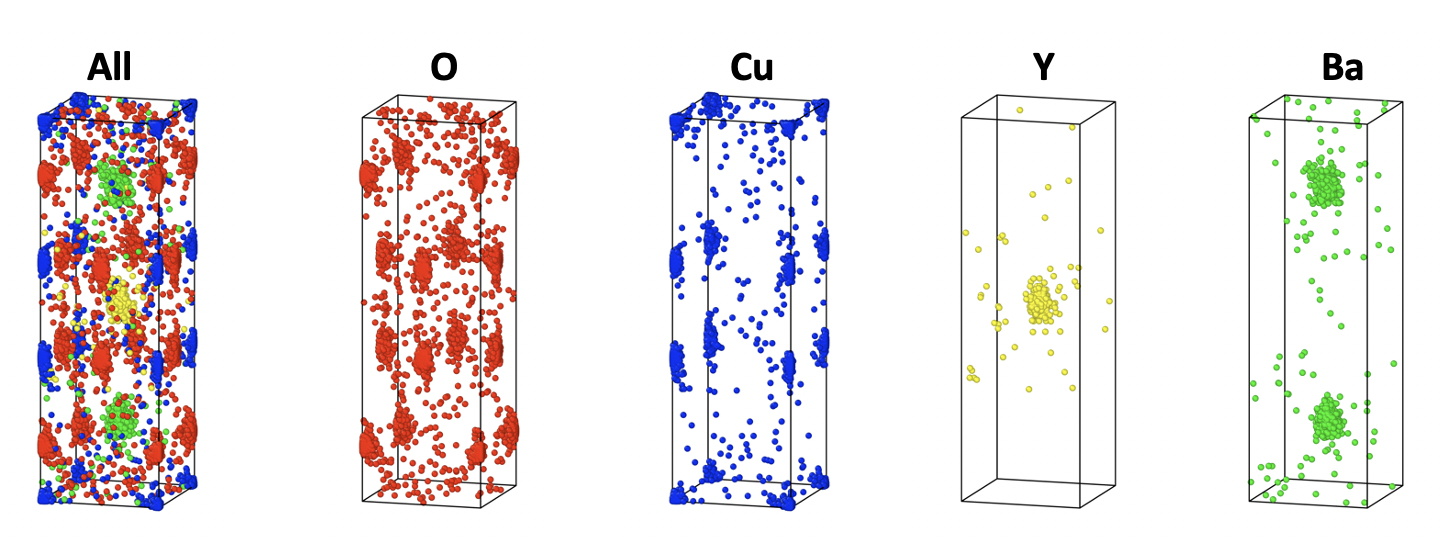}
    \caption{Cascade-damaged supercell folded into unit cells.}
    \label{folded}
\end{figure}

In the above figure, 177 antisite defects have been identified. Clearly, most of these defects arise from oxygen disorder in the material (in the amorphous region of the cascade). In fact, very few cation antisite defects actually appear to be present. As such, we can assume that the majority of the defects identified as antisites by Wigner-seitz analysis are in fact amorphous regions in cascade cores. 

\section{Oxygen Recombination in Sub-Stoichiometric YBCO}

\begin{figure}[h!]
    \centering
    \includegraphics[width=0.5\linewidth]{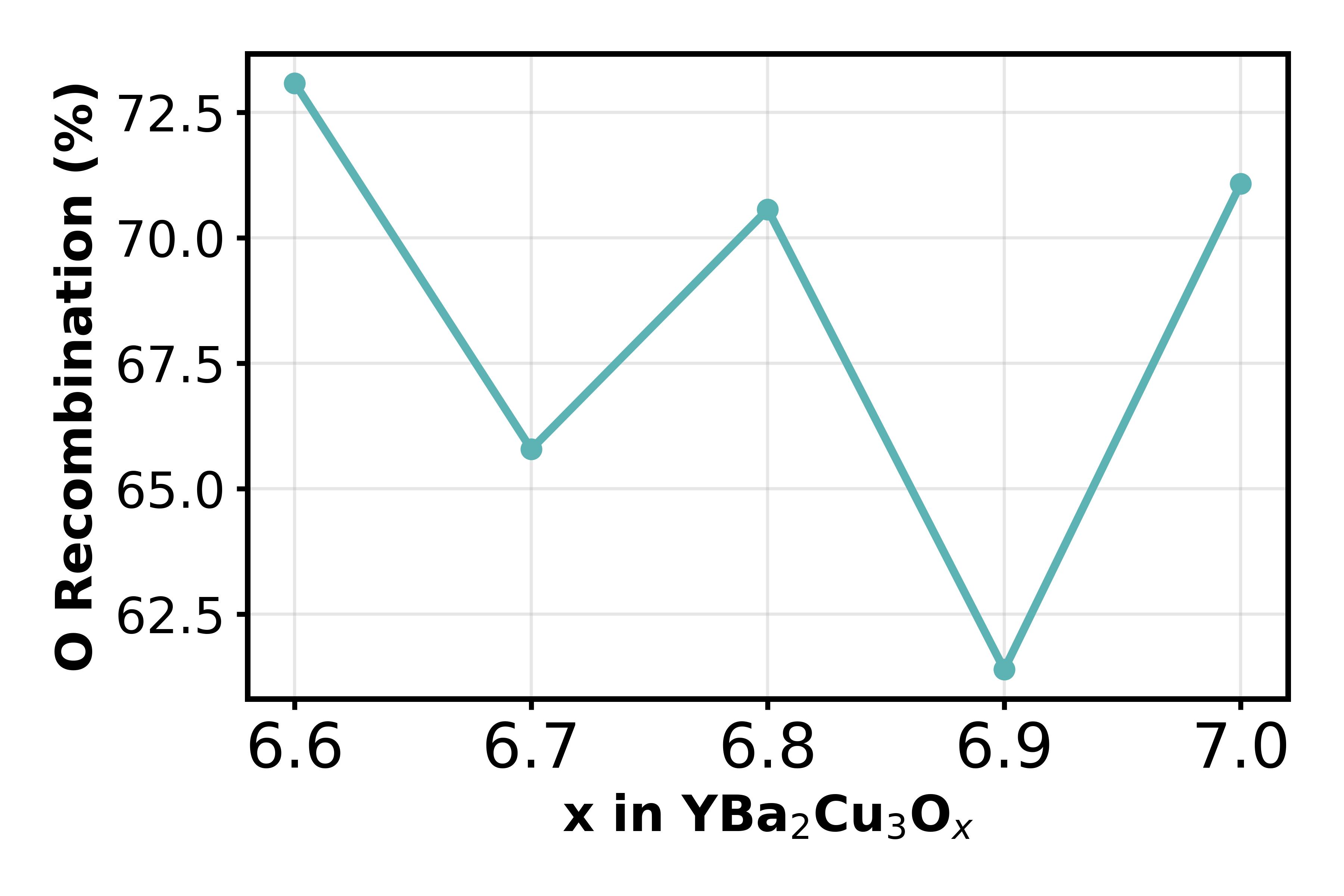}
    \caption{Mean oxygen recombination percentages during 5 keV collision cascades for different oxygen deficient YBCO cells.}
    \label{fig:Orecomb}
\end{figure}

Supplementary figure \ref{fig:Orecomb} demonstrates a potential weak increase in recombination percentage of oxygen with increasing oxygen deficiency. The noise in the data is significant due to the chaotic nature of cascades. Presumably, the oxygen vacancies found in the CuO chains for sub-stoichiometric YBCO aid diffusion of oxygen, enhancing the recombination percentage. Note that this is the recombination exhibited after 8 ps of simulation time. 

\section{Effect of Irradiation Direction on Defect Counts}

\begin{figure*}
    \centering
    \includegraphics[width=\linewidth]{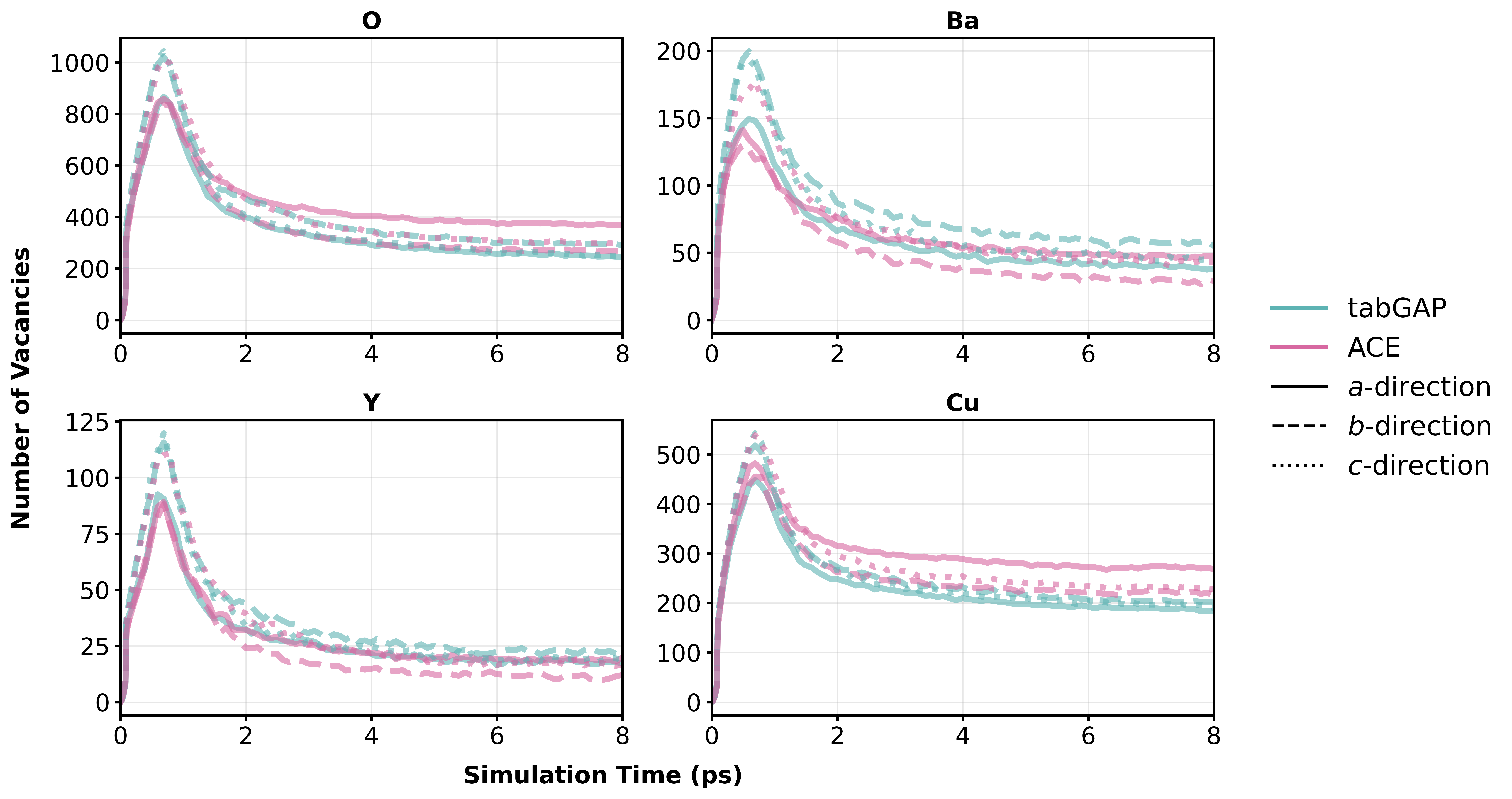}
    \caption{Number of vacancies of each element type for 5 keV Ba PKA collision cascades as a function of time. The solid line is the mean of all cascades for a given Cartesian direction. }
    \label{defsbydir}
\end{figure*}

Gray \textit{et al.} concluded that there was a weak directional dependence on the number of defects formed from cascades \cite{gray2022molecular}. To see if our MLPs reproduce the same behaviour, we have plotted the average vacancy counts for each set of cascades (where we performed 10 cascades in each Cartesian direction for each MLP). The results are shown in Figure \ref{defsbydir}. Interestingly, the peak defect counts in the $a$ and $b$ directions for the ACE are roughly equal. Contrastingly, the peak defect counts in the $b$ and $c$ directions are roughly equal for the tabGAP potential. This differing behaviour at the heat spike may have implications for the morphology of the damage region. However, the final defect counts do not show large differences by PKA direction.

\section{Potential Formalisms}

Machine learned potentials utilise the idea of a site energy ($\epsilon$), whereby the total quantum mechanical energy ($E$) of a system is split into per-site contributions:

\begin{equation}
    E = \sum_i \epsilon_i (\mathbf{d})
\end{equation}

\noindent{Here, $i$ represents the atom index, and $\mathbf{d}$ is an environment descriptor. Descriptors are mathematical encodings of structural information, which include invariance to permutation (exchange of like atoms), rotation and translation. The goal is to fit an MLP to the total energy ($E$) given the site energies ($\epsilon$) of the atoms in the system, described themselves by descriptors, $\mathbf{d}$. Derivative quantities (forces and sometimes stresses) are also invariably included in training, although for brevity we only include the details for fitting to energy below. The following section details the two formalisms of MLP compared in this paper.}

\subsection{GAP}

Here, we describe the GAP formalism (from the function space view of Gaussian-Process Regression \cite{deringer2021gaussian}), as one must first train a GAP before tabulating it to form the tabGAP. GAP describes the total energy of a system with a user-defined combination of descriptors. In our case, two-body, Embedded Atom Method (EAM) and three body descriptors are chosen. This breaks $E$ down into the following contributions:

\begin{equation}
    E = \sum_{p \in \mathsf{pairs}} \epsilon_p^{(2)} + \sum_{t \in \mathsf{triplets}} \epsilon_t^{(3)} + \sum_{m \in \mathsf{pairs}} \epsilon_m^{(eam)}
\end{equation}

\noindent{For pairs, triplets, and EAM pair densities (included in GAP in \cite{byggmastar2022multiscale}). However, since the individual site energies $\epsilon$ are not accessible from DFT calculations, we instead consider the covariance between the total energies of two configurations, $N$ and $M$:}

\begin{equation}
\langle E_N E_M \rangle = \left\langle \sum_{i \in N} \epsilon(\mathbf{d}_i) \sum_{j \in M} \epsilon(\mathbf{d}_j) \right\rangle = \sigma_w^2 \sum_{i \in N} \sum_{j \in M} C(\mathbf{d}_i, \mathbf{d}_j)
\end{equation}

This expression shows that by modelling the site energy, $\epsilon(\mathbf{d})$, as a function drawn from a Gaussian Process (GP), the total energy, $E$, becomes a linear combination of those GP function values. As a result, the covariance of total energies across configurations can be written entirely in terms of the kernel $C(\mathbf{d}_i, \mathbf{d}_j)$; a measure of similarity between local atomic environments.\\

The splitting of $E$ into site energies in this way enables a local, environment-based description of energy. The kernel $C$ is typically chosen as a squared exponential (Gaussian) kernel, and $\sigma_w$ is a descriptor specific scaling hyperparameter. These scalings are introduced to balance the contributions of different descriptor types (e.g. giving greater weight to two-body interactions than to higher-order terms, which are typically noisier or less constrained). \\

 Computation of the covariance function ($\mathbf{C}=\langle \mathbf{t}\mathbf{t}^T\rangle$), for past total energy observations ($\mathbf{t}$) for each descriptor allows one to estimate the mean predicted value ($\bar{y}$, the predicted total energy) of the underlying function at the next observation, given the previous $\mathbf{t}$ observations:

\begin{equation}
    \bar{y} = \mathbf{k}^T\mathbf{C}^{-1}\mathbf{t},
    \label{GPR}
 \end{equation}

\noindent{where $\mathbf{k}^T$ represents the covariance vector of function values ($\langle y\mathbf{t} \rangle$, for scalar function value $y$ at the new test input). Construction of the covariance matrices allows for the prediction of $\bar{y}$ for any arbitrary new point $t$. In the GAP formalism, this means that a potential energy surface can be constructed from $\mathbf{C}$, given the training data, for any arbitrary configuration. This shows the power of the GAP formalism: no original basis functions or weights are actually required; predictions depend entirely on the chosen kernel function $C$ and previous observations, $\mathbf{t}$. Note that in practice, computation of $\mathbf{C}$ is very expensive for large numbers of data points. Therefore, the model is trained on a sparsified set of training points. For more information see \cite{deringer2021gaussian}.}   \\

For the purposes of radiation damage, it is important to include a reliable short-range repulsive interatomic potential. The included repulsive potential ($V_{rep}$) is set to vanish at a point just below the equilibirum bond distance, so all near-equilibrium and long range interactions are fully machine learned. During fitting, $V_{rep}$ is calculated for all configurations in the DFT database, and is then subtracted from the DFT energies. The GAP is then fit as normal, and the energies are added back on at the end \cite{byggmastar2019machine}. This ensures a smooth fit between the two regimes. We use short ranged potentials fit to all-electron DFT data from \cite{nordlund2025repulsive}. \\

\subsection{tabGAP}

Evaluating equation \ref{GPR} is computationally expensive, resulting in relatively slow MLPs. This is where the utility of training on low dimensional descriptors comes in: the kernel can be tabulated, creating a ``tabGAP''. This was first introduced in \cite{glielmo2018efficient}, and later improved upon by \cite{byggmastar2021modeling}. Pairwise terms are evaluated as one dimensional interpolations between pair energies. Similarly, three body terms are tabulated as three-dimensional interpolations between triplet energies. This interpolation is performed on a grid of
($r_{ij}, r_{jk}, \cos{\theta_{ijk}}$) points. The EAM energies are tabulated as a standard EAM potential file, with 1D splines for the pair density function and embedding energies \cite{Byggmstar2022}. The total energy of the tabGAP is then: 

\begin{equation}
 E^{\mathrm{tabGAP}}_{\mathrm{tot}} = 
\sum_{i<j}^N S_{ij}^{1D}(r_{ij}) +
\sum_{i,j<k}^N S^{3D}_{ijk}(r_{ij},r_{ik}, \cos{\theta_{ijk}}) + \sum^N_i S^{1D}_{emb.}(\sum^{N}_j S_\phi (r_{ij})) .
\end{equation}

\noindent{where $S^{1D}$ and $S^{3D}$ represent 1D and 3D cubic spline interpolations, respectively. The repulsive pair potential and GAP pair potential are merged into one 1D spline, per element pair. The speed-up afforded can be several orders of magnitude, depending on the cutoffs used in the descriptors, and the types of descriptors used. }

\subsection{ACE}

Here, we describe the ACE formalism \cite{drautz2019atomic}. The configuration of the $i^{th}$ atom is describe as:

\begin{equation}
    \sigma_i = (r_i, \theta_i, h_i),
\end{equation}

\noindent{where $\sigma$ is a configuration with a position $r$, attributes $\theta$ (e.g. element), and learnable features $h$ (e.g. dipole moment). ACE is a systematic framework that allows for construction of symmetric polynomial representations of our input configurations $\sigma$. Starting with a one particle basis:}

\begin{equation}
    \phi_{v,z_i,z_j}(r) = R_{nl,z_i,z_j} r_{ji} Y_l^m(\hat{r})
\end{equation}

\noindent{where $i$ and $j$ are atom indices, $z$ is the atomic species, $\hat{r}$ is the unit vector for $r_{ij}$ (the distance between $i$ and $j$), $Y$ and $R$ are spherical and radial harmonics, and $v=nlm$ (quantum numbers corresponding to the harmonic basis). Translational invariance is built in by using absolute particle distances, and permutational invariance is built in as follows:}

\begin{equation}
    A_{i,v} = \sum_j \phi_v (\sigma_j, \sigma_i)
\end{equation}

\noindent{where the $A$ basis is evaluated for each neighbour and summed. Summation of  body-order expansion terms gives the total energy:}

\begin{equation}
\begin{split}
    E_i = V^{(1)}(r_i) + \dfrac{1}{2} \sum_j V^{(2)}(r_i,r_j) \\
    + \dfrac{1}{3!} \sum_{jk} V^{(3)}(r_i,r_j, r_k) + ...
\end{split}
\end{equation}

\noindent{for potential $V^N$ where $N$ represents the total $N^{th}$ body order. The $A$-basis is then used to parameterise site energy to yield:}

\begin{equation}
\begin{split}
    E_i = \overbrace{E_i^{(0)}}^\text{1 body} + \overbrace{\sum_v \tilde{c}_v^{(1)} A_{i,v}}^\text{2 body} + \overbrace{\sum_{v_1,v_2}^{v_1 \geq v_2} \tilde{c}_{v_1,v_2}^{(2)} A_{i,v_1} A_{i, v_2}}^\text{3 body} \\ + \overbrace{\sum_{v_1,v_2, v_3}^{v_1 \geq v_2 \geq v_3} \tilde{c}_{v_1,v_2,v_3}^{(3)} A_{i,v_1} A_{i, v_2}, A_{i, v_3}}^\text{4 body} + ...
    \label{ACEmanybody}
\end{split}
\end{equation}

\noindent{for expansion coefficients $\tilde{c}$. This can be generalised into the product:}

\begin{equation}
    \mathbf{A}_{i, \mathbf{v}} = \prod_{\xi=1}^v A_{i,v_{\xi}}
\end{equation}

This expression now needs to be symmetrised with respect to rotations $Q$. This can be written mathematically as:

\begin{equation}
    \mathbf{B}_{i,\mathbf{v}} = \int_{O(3)} \mathbf{A}_{i,\mathbf{v}}(\lbrace Q.(\sigma_i, \sigma_j) \rbrace_{j\in \mathcal{N}(i)}) dQ
\end{equation}
\noindent{where $\mathcal{N}(i)$ denotes atoms neighbouring atom $i$. $O(3)$ in this case is the group of all orthogonal $3\times 3$ matrices that represent transformations preserving angles and distances. Effectively, the expression averages over all possible rotations in the system. This can be more simply expressed as a product in the $A$-basis using Clebsch-Gordan coefficients}

\begin{equation}
    \mathbf{B}_{i,\mathbf{v}} = \sum_{\mathbf{w}} C_{\mathbf{v},\mathbf{w}} \mathbf{A}_{i,\mathbf{v}}
\end{equation}

Therefore, the complete linear ACE energy expression is:

\begin{equation}
    E = \sum_{\mathbf{i}, \mathbf{v}} c_{i,\mathbf{v}} \mathbf{B}_{i,\mathbf{v}}
\end{equation}

 The ACE model is then trained by minimising a loss function of the total energy, forces and virials/stresses.






\end{document}